\documentclass[10pt,conference]{IEEEtran} 
\IEEEoverridecommandlockouts
\usepackage{cite}
\usepackage{amsmath,amssymb,amsfonts}
\usepackage{algorithmic}
\usepackage{graphicx}
\usepackage{textcomp}
\usepackage{xcolor}

\usepackage{subcaption}
\usepackage{verbatim}
\usepackage{multirow}
\usepackage{epigraph}
\usepackage{tcolorbox}
\usepackage{enumitem}
\usepackage{booktabs}
\usepackage{pifont}
\usepackage{url}
\usepackage{hyperref}
\renewcommand{\tablename}{Table}
\renewcommand{\thetable}{\arabic{table}}
\usepackage{adjustbox}
\definecolor{LightGray}{gray}{0.9}

\definecolor{keyword}{RGB}{255,0,90}
\definecolor{comment}{RGB}{0,128,0}
\definecolor{string}{RGB}{0,0,255}

\def\BibTeX{{\rm B\kern-.05em{\sc i\kern-.025em b}\kern-.08em
    T\kern-.1667em\lower.7ex\hbox{E}\kern-.125emX}}

\newcommand{\major}[1]{{\color{black}#1}}
    
\begin{document}


\title{Knowledge-Enhanced Program Repair for \\Data Science Code \vspace{-0.3em}}


\author{
\IEEEauthorblockN{Shuyin Ouyang}
\IEEEauthorblockA{\textit{King's College London} \\
London, UK \\
shuyin.ouyang@kcl.ac.uk}
\and
\IEEEauthorblockN{Jie M. Zhang}
\IEEEauthorblockA{\textit{King's College London} \\
London, UK \\
jie.zhang@kcl.ac.uk}
\and
\IEEEauthorblockN{Zeyu Sun}
\IEEEauthorblockA{\textit{Institute of Software, Chinese} \\
\textit{Academy of Sciences}, Beijing, China\\
zeyu.zys@gmail.com}
\and
\IEEEauthorblockN{Albert Merono Penuela}
\IEEEauthorblockA{\textit{King's College London} \\
London, UK \\
albert.merono@kcl.ac.uk}
}

\IEEEaftertitletext{\vspace{-2.5em}}
\maketitle

\thispagestyle{plain}
\pagestyle{plain}

\begin{abstract}

This paper introduces DSrepair, a knowledge-enhanced program repair approach designed to repair the buggy code generated by LLMs in the data science domain.
DSrepair uses knowledge graph based RAG for API knowledge retrieval and bug knowledge enrichment to construct repair prompts for LLMs.
Specifically, to enable knowledge graph-based API retrieval, we construct DS-KG (Data Science Knowledge Graph) for widely used data science libraries.
For bug knowledge enrichment, we employ an abstract syntax tree (AST) to localize errors at the AST node level. 
\major{We evaluate DSrepair’s effectiveness against five state-of-the-art LLM-based repair baselines using four advanced LLMs on the DS-1000 dataset.}
\major{The results show that DSrepair outperforms all five baselines.}
Specifically, when compared to the second-best baseline, 
\major{DSrepair achieves substantial improvements,}
fixing 44.4\%, 14.2\%, 20.6\%, and 32.1\% more buggy code snippets for each of the four evaluated LLMs, respectively.
Additionally, it achieves greater efficiency, reducing the number of tokens required per code task by 17.49\%, 34.24\%, 24.71\%, and 17.59\%, respectively.

\end{abstract}


\section{Introduction}

Data science is crucial in driving innovation and decision-making across various domains~\cite{bolyen2019reproducible, hassani2023role}, leveraging data to uncover insights and inform strategic actions. 
Nevertheless, 
\major{the complexity of data science libraries and the expertise required to use them} can pose significant barriers to lay users. 
Large Language Models (LLMs) have emerged as powerful tools to generate data science code automatically~\cite{hong2024data, nejjar2023llms, lai2023ds}, democratizing access and accelerating development processes.
Despite their potential, the widely acknowledged shortcomings with LLMs, such as hallucination and the lack of specialized knowledge of certain domains (e.g. long-tail API usage)~\cite{zan2023private, zan2022language} and specific code context~\cite{ge2024openagi}, remain significant obstacles. 
These issues are particularly critical in the data science domain, where the code heavily relies on libraries for accurate and efficient data processing and analysis, making 
precision and contextual accuracy essential for robust outcomes.
Existing studies have applied feedback-based iterative self repair~\cite{xia2023keep, gupta2020synthesize, chen2023teaching, fu2023chatgpt, zhang2023critical}
to improve the reliability of LLM-generated code.
Nevertheless, these \major{approaches} are not designed for data science code.

Recently, Retrieval-Augmented Generation (RAG)~\cite{lewis2020retrieval} has also emerged as a widely-adopted technique to inject external knowledge into LLMs to facilitate more coherent code generation.
RAG combines the strengths of information retrieval and LLMs to enhance code generation.
\major{Existing RAG-based code generation studies~\cite{parvez2021retrieval, li2023acecoder, zhou2022docprompting, liu2023codegen4libs} commonly follow a standard RAG architecture,}
where the ``retriever'' component retrieves relevant \textbf{plain text} from a vast corpus or database using a vector similarity search. This retrieved information is then fed into an LLM, which uses this context to produce more accurate and relevant code\cite{ahmed2024automatic, lu2022reacc, tang2023domain, zhang2023repocoder}.
\major{However, these text-based RAG approaches are not well-suited for code generation tasks, as they rely on unstructured or semi-structured plain text, which lacks the semantic relationships and structured representation needed for complex code understanding and generation.}
Specifically,
1) text-based retrieval relies on vector similarity search, which often retrieves irrelevant or loosely related information due to ambiguities in natural language;
2) plain text does not explicitly represent the relationships between APIs, their dependencies, or their attributes (e.g., parameters, return types). As a result, text-based RAG \major{approaches} may fail to provide the comprehensive contextual knowledge required for resolving an issue;
3) retrieved plain text often contains redundant descriptions or ambiguities, which can confuse large language models (LLMs) or lead to suboptimal code generation.

This paper introduces DSrepair, a knowledge-enhanced \major{approach} for repairing incorrect data science code produced by LLMs through knowledge graph-based RAG and bug information enrichment. 
We construct  DS-KG (\textbf{D}ata \textbf{S}cience \textbf{K}nowledge \textbf{G}raph),
a set of knowledge graphs for the seven most widely adopted data science libraries (i.e., NumPy, Pandas, SciPy, Scikit-learn, Matplotlib, PyTorch, and TensorFlow)~\cite{yin2018learning, yin2022natural, lai2023ds, hong2024data, nejjar2023llms}.
For the buggy code generated by an LLM, 
DSrepair uses the API name appeared in the code as the query,
to obtain the correct usage of corresponding API functions by accessing DS-KG.
It then uses the returned result to guide the LLM in repairing code.

Compared to text-based RAG, our KG-based RAG naturally captures more complex relationships and dependencies within API documents, which is essential in the data science domain.
\major{The rich semantic relationships stored in knowledge graphs enhance the reliability and efficiency of code generation and repair.}
For instance, in Matplotlib's API document, an API named \textit{matplotlib.pyplot.subplots}, has a parameter called \textit{gridspec\_kw}.
The value of \textit{gridspec\_kw} must be passed to another API object, called \textit{matplotlib.gridspec.GridSpec}.
\major{If an error occurs in \textit{gridspec\_kw}, querying knowledge from \textit{matplotlib.gridspec.GridSpec} is more helpful than querying only from \textit{matplotlib.pyplot.subplots}.
}
A well-designed KG can infer such dependency naturally.




DSrepair also uses enhanced bug information to improve the program repair effectiveness.
Data science code often contains multiple function calls and data operations within a single line.
Therefore, to obtain fine-grained bug information, DSrepair 
uses Abstract Syntax Tree (AST) and test case execution information to localize errors at the AST node level.

We evaluate DSrepair on two widely used general-purpose LLM (i.e., GPT-3.5-turbo and GPT-4o-mini~\cite{gptmodel}) and two state-of-the-art coding LLMs (i.e., DeepSeek-Coder~\cite{guo2024deepseek} and Codestral~\cite{codestral}) based on DS-1000~\cite{lai2023ds}, the data science code generation benchmark spanning seven data science libraries.
Our results show that DSrepair outperforms all the five baseline LLM-based repairing \major{approaches}.
In particular, compared to the second-best baseline,
DSrepair correctly fixes 44.4\%, 14.2\%, 20.6\%, and 32.1\% more buggy code snippets for the four LLMs, respectively,
\major{while reducing token usage per code task by 17.49\%, 34.24\%, 24.71\%, and 17.59\%.}

To summarize, this paper makes the following contributions:

\begin{itemize}[leftmargin=*]
    \item  We present DSrepair, a novel LLM-based program repair \major{approach} for data science code, leveraging knowledge graph-based RAG and enriched bug information. 
    \item We construct and release DS-KG (Data Science Knowledge Graph), a comprehensive set of knowledge graphs tailored to the seven most widely used data science libraries.  
    \item We propose an AST-based bug information enriching \major{approach} that can pinpoint errors at the AST node level.
    \item We conduct an empirical study using four LLMs and five baselines, demonstrating that DSrepair significantly outperforms all baselines in repairing data science code.
\end{itemize}

We release our data, code, KG data dump, and results at our homepage~\cite{homepage}.
The rest of the paper is organized as follows.
Section~\ref{section: method} outlines our methodology.
Section~\ref{section: experimental design} describes the design of the experiments, including research questions, benchmarks, baselines, selected models, and measurements.
Section~\ref{section: result and findings} presents the results and highlights notable findings based on our empirical results.
\major{Section~\ref{section: discussion} discusses the threat to validity, the limitation, and the generalizability of our work.}
Section~\ref{section: related work} introduces the related work of our study.
Section~\ref{section: conclusion} concludes.

\section{Method}
\label{section: method}

\begin{figure}[t]
\centerline{\includegraphics[width=0.89\linewidth]{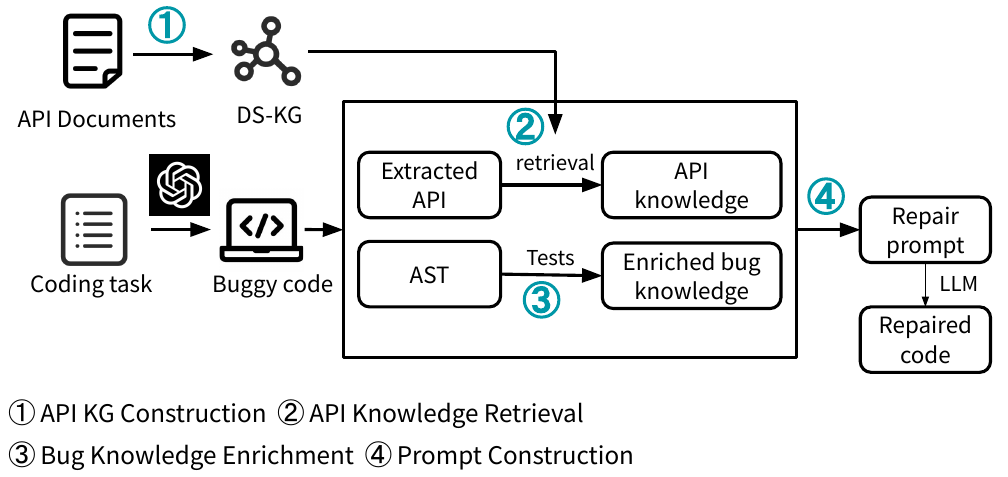}}
\vspace{0mm}
\caption{Overview of DSrepair.}
\label{fig: overview}
\vspace*{-0.4cm}
\end{figure}

Fig~\ref{fig: overview} shows an overview of DSrepair.
Given a code problem description, we first let LLM (i.e. GPT-3.5-turbo) generate code.
If the code fails to pass the test cases, 
DSrepair constructs a repair prompt \major{and requests} LLM to regenerate the code (see Section~\ref{sec:benchmark} for details).
As shown in the figure, DSrepair involves four main steps: API KG Construction, where a knowledge graph (DS-KG) is built for popular data science libraries (e.g., NumPy and Pandas) to capture detailed API usage and relationships; API Knowledge Retrieval, where API calls are extracted from the buggy code and queried from DS-KG, with the results verbalized into natural language for LLM prompts; Bug Knowledge Enrichment, which localizes errors at the AST node level using test case execution to provide fine-grained bug information; and Prompt Construction, where all gathered information is structured into a detailed prompt to guide the LLM in generating effective repairs.


\subsection{API KG Construction}

\begin{figure*}[t]
\centerline{\includegraphics[width=0.9\linewidth]{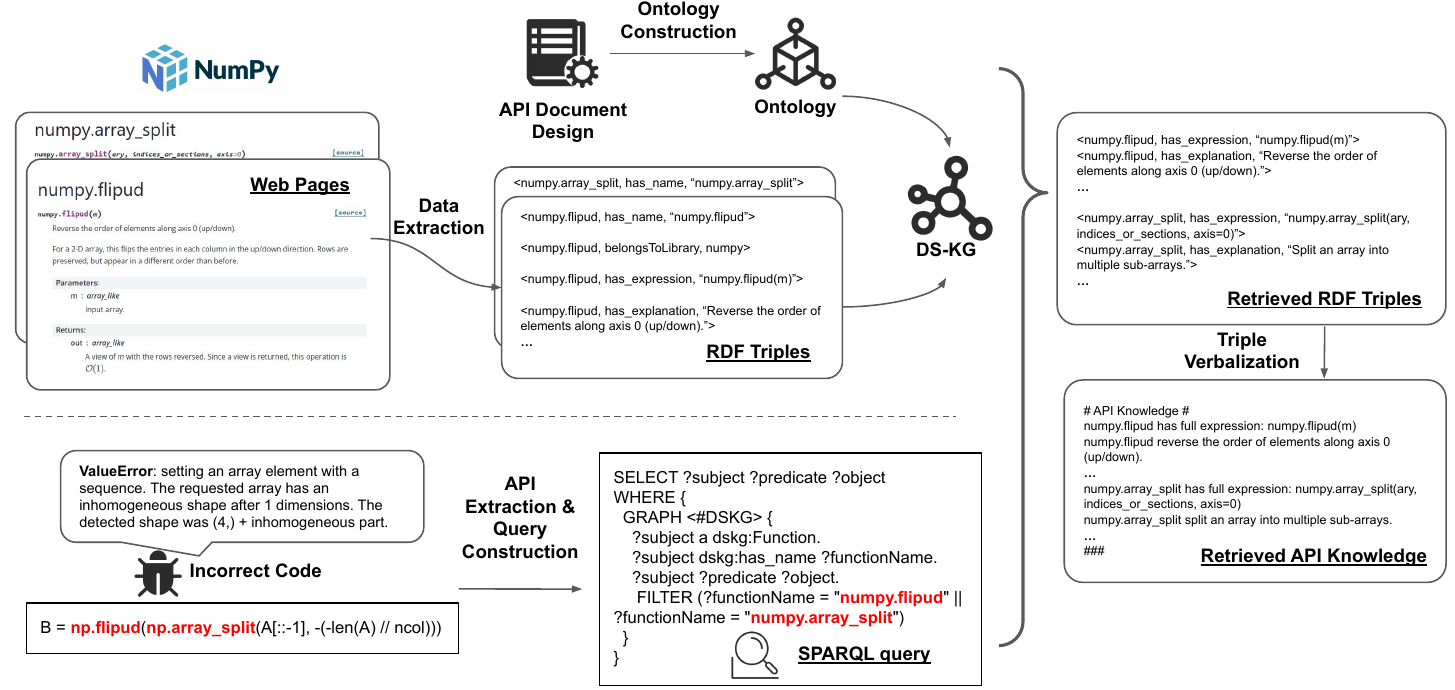}}
\vspace{0mm}
\caption{
Details on API KG construction (Step 1) and API Knowledge Retrieval (Step 2) in DSrepair. 
The code raises an error because of the mismatched array shape between \textit{np.flipud}'s required input and \textit{np.array\_split}'s output.
DSrepair extracts the API call in the error line and builds a SPARQL query to search the relevant RDF triples in the DS-KG, which is constructed from API documents and guided by the ontology.
Finally, DSrepair maps the returned RDF triples to natural language, which will be used as a part of the repair prompt.
}
\label{fig: DSKG example}
\vspace*{-0.4cm}
\end{figure*}

We develop DS-KG, a knowledge graph tailored to widely adopted data science libraries such as NumPy, Pandas, SciPy, Scikit-learn, Matplotlib, PyTorch, and TensorFlow.
Its primary purpose is to assist LLMs in repairing buggy code by providing structured information about the correct usage of APIs.
Following standard KG construction procedures, we begin by creating an ontology to define the schema for DS-KG~\cite{simperl2014collaborative, suarez2011neon}.
Existing ontologies are unsuitable for representing API documentation in the context of data science code repair.
To address this gap, we manually design a domain-specific ontology schema, drawing insights from the structure of API documentation.
API documentation typically provides details such as an API’s name, expression, explanation, parameters, and return types.
Our ontology captures these attributes for individual API functions, enabling precise and structured queries based on error information extracted from buggy code.
Inspired by prior work in code ontology design~\cite{liang2022misusehint, abdelaziz2021toolkit}, we represent each API function as a unique entity within DS-KG.
The ontology includes two types of relations:
(1) Attribute Relations, describe links between API entities and their attributes, such as: `\textit{has\_name}', `\textit{has\_expression}', and `\textit{has\_explanation}'\footnote{In this paper, we ignore the OWL prefixes in RDF triples (\textless subject, predicate, object\textgreater) to make the article more concise.}.
(2) Dependency Relations, capture the hierarchical structure and dependencies of APIs, such as `\textit{belongsToLibrary}' and `\textit{belongsToModule}'.

Fig~\ref{fig: DSKG example} illustrates an example of DS-KG construction from the NumPy API document.
Each API object introduced on a webpage, such as \textit{numpy.flipud} and \textit{numpy.array\_split}, is treated as an entity.
Detailed information about an API object, such as its name, expression, explanation, and parameters \& returns, is used to build RDF triples with attribute relations.
For example, the API object \textit{numpy.flipud} has such an RDF triple, \textless \textit{numpy.flipud}, \textit{has\_expression}, ``\textit{numpy.flipud(m)}''\textgreater.
New entities are created for the parameters and return values of each API object, each with attributes like argument position, data type, and explanation.
For example, the parameter \textit{m} in ``\textit{numpy.flipud(m)}'' has the following RDF triple: \textless \textit{numpy.flipud\_parameter\_m}, \textit{hasType}, ``\textit{array\_like}''\textgreater.
Using the prefix of the API entity (derived from the name and webpage URL), we construct RDF triples with dependency relations.
For instance, the API object \textit{numpy.flipud} is linked to its library through the RDF triple: \textless \textit{numpy.flipud}, \textit{belongsToLibrary}, \textit{numpy}\textgreater.

\subsection{API Knowledge Retrieval}

DSrepair integrates DS-KG to enhance the repair of buggy code by retrieving relevant API knowledge and incorporating it into the repair process.
DSrepair extracts all API invocations in the buggy code snippet using regular expressions (e.g., identifying `\textit{np.flipud}' or `\textit{np.array\_split}'). 
It resolves API prefixes (e.g., mapping `\textit{np}' to `\textit{numpy}') and uses the full API name for queries, accounting for the common use of abbreviations in data science libraries.
Using the resolved API name, DSrepair constructs SPARQL queries~\cite{prudhommeaux2008sparql} to retrieve RDF triples from DS-KG.
These triples encapsulate knowledge specific to the queried API, such as its attributes, dependencies, and parameter details.
To ensure compatibility with LLMs, we transform the retrieved RDF triples into natural language sentences using triple verbalization techniques~\cite{blinov2020semantic}.
These sentences provide human-readable explanations, including a description of the API’s purpose and syntax, details about parameters and returns together with their data types and explanations.
The retrieved API knowledge is concatenated and included in the ``API Knowledge'' section of the repair prompt provided to the LLM.

In Section~\ref{section: API richness}, we demonstrate that incorporating only the full API expression as knowledge yields the best performance for data science code repair.
Thus, by default, DSrepair includes the full API expression in the prompt.
This approach balances the richness of information and efficiency, ensuring LLMs receive sufficient contextual guidance without overwhelming them with unnecessary details.

\begin{figure*}[htbp]
\centerline{\includegraphics[width=0.8\linewidth]{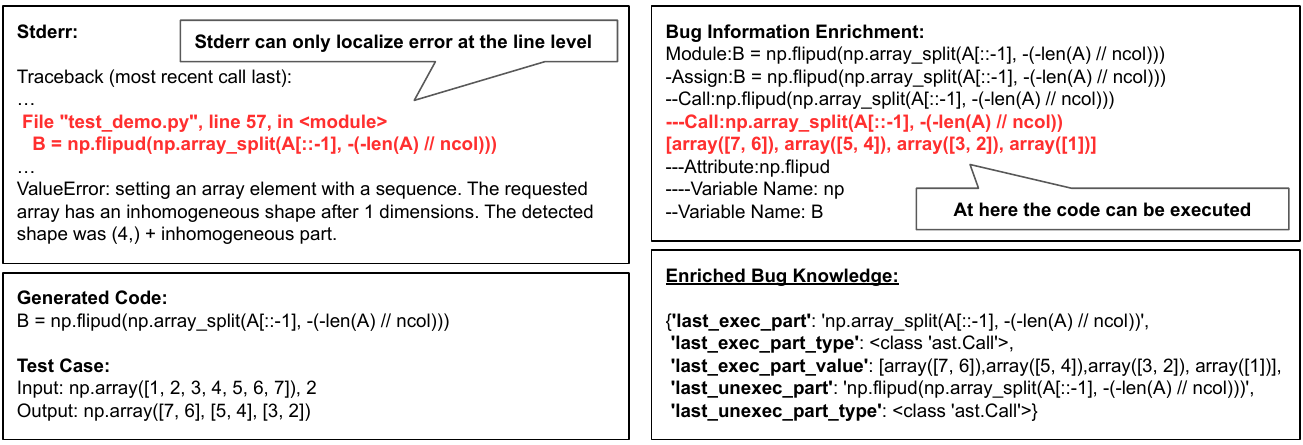}}
\vspace{0mm}
\caption{Bug knowledge enrichment example. Stderr (standard error information) can only localize the bug at the line level, while our bug knowledge enrichment could enrich the error information to the AST node level.}
\label{fig: AST_fault_localization}
\vspace*{-0.5cm}
\end{figure*}

\subsection{Bug Knowledge Enrichment}
\label{sec: bug knowledge enrichment}

Bug knowledge enrichment aims to
provide LLMs with extra bug information to help LLMs better repair the bug without requesting extra tests.
We use only the example tests provided in the coding task description.
Traditional fault localization \major{approaches} such as spectrum-based fault localization~\cite{abreu2007accuracy} and mutation-based fault localization~\cite{papadakis2015metallaxis} are not applicable here for two reasons.
First, data science code generation benchmarks usually provide a very limited number of tests (e.g., 1.6 tests on average per problem in DS-1000) since the annotators need to define program inputs with complex objects such as square matrices, classifiers, or dataframes~\cite{lai2023ds};
second, traditional \major{approaches} are often file-level or line-level fault localization, while data science code tends to
contain multiple function calls and data operations in one line. 
Therefore, different from traditional \major{approaches}, DSrepair uses AST-node level bug information to provide LLMs with more fine-grained bug information.
Fig.~\ref{fig: AST_fault_localization}
shows a specific example of our bug information enrichment procedure.

Firstly, test cases are extracted from the coding task description provided.
These tests are essential for validating the correctness of the code and are used later in the bug knowledge enrichment process.
We then 
transform the incorrect code snippet into its AST representation.
Once the AST is generated, DSrepair iterates within a namespace that includes all necessary libraries and the extracted test cases.
This iteration involves executing nodes in the AST sequentially.

To gain detailed bug information, the system identifies the last unexecuted node in the AST.
We classify all the bugs into two categories:
1) Runtime Errors.
If the code contains bugs that prevent it from being executed, the system will run each AST node until it encounters an error.
The AST node that was executable before the failure occurred is noted as the last executed node.
The node immediately following this, which causes the failure, is where the bug is likely located.
The error is between these two nodes: the last executable node and the first unexecutable node.
2) Assertion Errors. 
If the entire code can be executed but the results do not match the expected output, such an issue can be due to an assertion error.
In this case, the system captures the final value returned by the code execution.
By comparing this actual output with the desired result, the system can provide information to LLMs about why the code is incorrect.
The comparison highlights discrepancies, offering insights into potential logical errors in the code.

\subsection{Prompt Construction}
\begin{figure}[t]
\centerline{\includegraphics[width=0.6\linewidth]{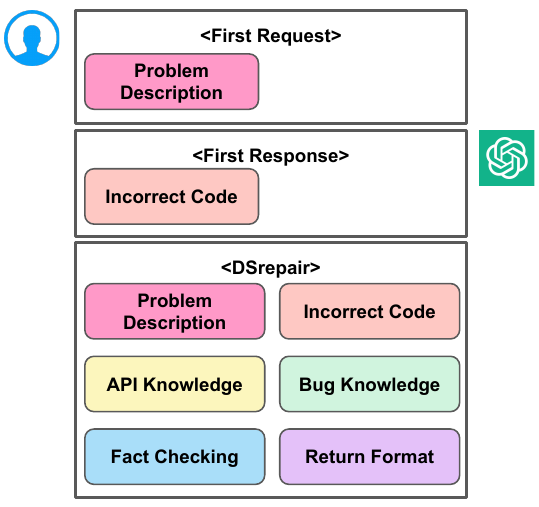}}
\vspace{0mm}
\caption{DSrepair prompt example. The prompt contains structural and rich information to guide LLMs for code generation.}
\label{fig: prompt example}
\vspace*{-0.6cm}
\end{figure}

This step uses information obtained from the previous steps and organizes it into a structured prompt~\cite{brate2022improving}, which is then fed to the LLM for code repair.
As shown in Fig~\ref{fig: prompt example},
the final prompt includes the following components: problem description, incorrect code, stderr information, API knowledge, bug knowledge, fact-checking, and response format.

We first put the problem description and LLM generated incorrect code in the prompt.
Error messages are cleaned by removing local file paths and deleting warnings.
To some extent, this action protects the privacy of users' operation system environment and ensures that only relevant error information is included, focusing on critical errors that hinder code execution.
 
We extract useful API knowledge from the DS-KG query results, specifically the API expression or signature.
This expression includes all parameters both compulsory and optional highlighting potential errors related to parameter usage and function calls.
This comprehensive parameter information is crucial as it often points to the source of errors in the code.

For bug knowledge, we leverage the results from bug enrichment.
This involves providing the test case, the last unexecutable AST node, and the last executable AST node along with the executed result value.
By comparing the actual output with the desired output, we can pinpoint the exact location and nature of the error.
This detailed local context helps in understanding the specific issues within the code.

The fact-checking component identifies where the existing code logic violates the corresponding requirements outlined in the problem description.
This step is essential to redirect the LLM's attention back to the problem description, ensuring that the solution aligns with the original requirements and constraints.
The complete prompts that we use are available on our homepage~\cite{homepage}.



\section{Experimental Design}
\label{section: experimental design}

\subsection{Research Questions}

Our evaluation answers the following questions:

\noindent \textbf{\emph{RQ1: How effective is DSrepair in repairing LLM-generated data science code?}} This RQ investigates the buggy code fix rate of DSrepair compared with other state-of-the-art program repair \major{approaches}.

\noindent \textbf{\emph{RQ2: How do DSrepair's bug fixes overlap with the five baselines?}}
This RQ investigates whether DSrepair could fix unique bugs that the baselines fail to address.

\noindent \textbf{\emph{RQ3: What is the cost of DSrepair?}} This RQ investigates token usage and money spent on DSrepair and our baselines.

\noindent \textbf{\emph{RQ4: How is DSrepair's performance affected by different prompt designs?}} To understand how different prompt designs affect DSrepair, we conduct an ablation study to analyze each key prompt component's contribution to DSrepair.

\noindent \textbf{\emph{RQ5: How do different knowledge retrieval approaches affect DSrepair?}} 
This RQ aims to explore the advantage of knowledge graph-based RAG against plain text-based RAG.

\noindent \textbf{\emph{RQ6: How does the richness of API knowledge affect DSrepair?}}
This RQ studies whether different types of API knowledge (e.g., whether the knowledge contains explanation or parameters) given in DSrepair will affect its performance.

\noindent \textbf{\emph{RQ7: How does the non-determinism of LLM affect our experiment results?}} This RQ studies the influence of LLM's inherent randomness on our experiment results.

\subsection{Data Science Benchmark}
\label{sec:benchmark}
Our evaluation uses DS-1000~\cite{lai2023ds},
the state-of-the-art benchmark specifically designed for benchmarking LLMs in data science code generation.
The DS-1000 benchmark was specifically constructed to mitigate concerns about data leakage. In particular, the dataset applies perturbation (e.g., text rephrasing and semantic perturbation),
so that models cannot answer them correctly by memorizing the solutions from pre-training~\cite{lai2023ds}. 

Other data science code generation benchmarks~\cite{agashe2019juice, yin2018learning} are not applicable because they are not based on realistic
problems and have no dedicated test cases to evaluate the correctness of the code (they use Exact Match~\cite{exact_match} or BLEU score~\cite{post2018call}).
DS-1000 comprises one thousand diverse and practical data science problems sourced from StackOverflow, covering seven essential Python libraries:
Numpy\cite{numpy}, Pandas\cite{pandas},
Scipy\cite{scipy}, Sklearn\cite{sklearn},
Matplotlib\cite{matplotlib},
PyTorch\cite{pytorch}, and TensorFlow\cite{TensorFlow}.
The version of each library can be found on our homepage~\cite{homepage}.

In our experiments,  we let GPT-3.5-turbo generates code for each of the 1000 coding tasks in DS-1000, 562 of the generated programs fail to pass the test cases, and are regarded as repair targets of DSrepair.

\subsection{Baseline}
As far as we know, there are no existing repair \major{approaches} that are specifically designed for data science code generation.
Therefore, in our evaluation, we compare DSrepair against the following state-of-the-art LLM-based \major{approaches} that are capable of repairing general types of code.
While these general-purpose \major{approaches} are effective in many scenarios, they are not tailored to address the unique challenges posed by data science-specific bugs. By addressing the distinct requirements of data science code, we aim to demonstrate DSrepair’s enhanced ability to handle data science-specific repair tasks in comparison to these baselines.

\textbf{Code-Search}~\cite{chen2024code}: Code-Search guides code repair by searching for similar code in the code base and adding the search result as a suggestion to the prompt. 
Following the practice in the paper, we use the code problem description as the query, and Lucene~\cite{lucene} as searching engine to conduct code search in the code base PyTorrent~\cite{bahrami2021pytorrent}.

\textbf{Chat-Repair}~\cite{xia2023keep}: Chat-Repair leverages the code execution result to check code correctness.
If the code cannot pass the test cases, Chat-Repair incorporates the execution results in the prompt, to provide richer information for code debugging.

\textbf{Self-Debugging-S}~\cite{chen2023teaching}: Self-Debugging-S (S represents Simple) enriches the prompt with the simplest information, a sentence that indicates the code's correctness without more detailed information.
For instance, ``The generated code is incorrect. Please fix the code.''

\textbf{Self-Debugging-E}~\cite{chen2023teaching}: Self-Debugging-E (E represents Explanation) first requests LLM to generate a line-by-line explanation about intermediate execution steps of the generated code.
Then, it requests LLM again to generate code, based on the line-by-line explanation of the incorrect code.

\textbf{Self-Repair}~\cite{olausson2023demystifying}: Self-Repair first leverages error information produced by test execution to make LLM produce a short explanation of why the code failed.
Then, it uses the explanation as part of the prompt to request LLM to improve the incorrectly generated code.

\subsection{LLMs}

We use two widely used general-purpose LLMs (GPT-3.5-turbo and GPT-4o-mini~\cite{gptmodel}) and 
two state-of-the-art coding LLMs (DeepSeek-Coder~\cite{guo2024deepseek} and Codestral~\cite{codestral})\footnote{We do not use GPT-4 because it comes at a significantly higher cost.}.
We access all the LLMs by using their commercial APIs.
The details of the four LLMs are shown in Table~\ref{tab: LLM detail}.
We choose Python as our programming language for the code generation tasks because DS-1000 is based on Python.


To control the randomness, 
we set the temperature of all the LLMs to 0.
For each \major{approach}, we let the LLMs generate code for each coding task ten times\footnote{We repeat experiments for Deepseek-Coder-V2 three times only, because the API is no longer available after 2024/09/05.}.
We select the result with median overall performance as the final result~\cite {ouyang2023llm}.
Our RQ7 in Section~\ref{section: RQ7} is about the influence of LLM's inherent randomness on our experiment results.

\begin{table}[htbp]\scriptsize
\caption{LLMs used in the evaluation.}
\vspace*{-0.4cm}
\begin{center}
\setlength{\tabcolsep}{4pt}
\begin{tabular}{l r r r}
\toprule
LLM & Version & Input Token Price & Output Token Price \\

\midrule
GPT-3.5-turbo & GPT-3.5-turbo-0125 & \$0.50/1M tokens & \$1.50/1M tokens \\
GPT-4o-mini & GPT-4o-mini-2024-07-18 & \$0.15/1M tokens & \$0.60/1M tokens \\
DeepSeek-Coder & DeepSeek-Coder-V2 & \$0.14/1M tokens & \$0.28/1M tokens \\
Codestral & Codestral-2405 & \$1.00/1M tokens & \$3.00/1M tokens \\
\bottomrule
\end{tabular}
\label{tab: LLM detail}
\end{center}
\vspace*{-0.7cm}
\end{table}

\subsection{Measurement}
\label{section: measurement}

We introduce the following metrics for measuring the performance of DSrepair and baselines.

\textbf{Effectiveness}: We measure the effectiveness of different approaches by checking their capability in fixing incorrectly generated code, including the Absolute Number of Fixes (ANF) and Fix Rate (FR).
The former is the absolute number of coding tasks whose code is successfully fixed. 
The latter is the ratio of ANF against all the buggy code snippets.
For ANF, two of the authors conduct manual verification on the correctness of the patches to make sure that the reported fixes are not overfitted.

\textbf{Cost}: We measure the cost by Token Usage (TU) and Money Spent (MS), which are the most widely used metrics for measuring cost for LLM-based approaches~\cite{xia2023keep}. 
TU refers to the total token usage when using LLM to finish one complete request on average, including input token usage and output token usage.
MS refers to the money cost for LLM to receive and return those tokens.
Below is the formula for the MS:
\vspace{-2mm}
\begin{equation*}
MS = \sum_{n=1}^{N} {(Token_{i,n} \times P_i + Token_{o,n} \times P_o)}
\end{equation*}
\vspace{-1mm}
where $P_i$ and $P_o$ refer to the input and output token price, $Token_{i,n}$ and $Token_{o,n}$ refer to the input token usage and output token usage at certain request $n$, and $N$ refers to the total number of LLM requests.

\section{Results}
\label{section: result and findings}

This section introduces the experimental results as well as the analysis and findings for each RQ.

\subsection{RQ1: Effectiveness of DSrepair}

To answer RQ1, we report the results of Absolute Number of Fixes (ANF) and Fix Rate (FR) for DSrepair and all the baselines with each LLM.
DSrepair initially generates 555 patches that successfully pass the tests from all the LLMs. After manual checking, two of the patches generated by GPT-4o-mini are overfitted\footnote{An overfitted patch passes the test cases but is actually incorrect.
} and have been removed from the repaired set.
Table~\ref{table: FR all library categories} shows the ultimate results.

We can observe that DSrepair significantly outperforms all the baselines in terms of ANF and FR across all four LLMs we study.
Specifically, 
DSrepair can fix the buggy code for 104, 145, 164, and 140 coding tasks for the four LLMs, respectively,
while the second-best results are 72, 127, 136, and 106, respectively.

For specific data science libraries, 
DSrepair outperforms the baselines for most libraries. 
For example, for GPT-3.5-turbo, DSrepair has the highest fix rate in Numpy, Scipy, Sklearn, Matplotlib, and PyTorch.
For Codestral, DSrepair performs the best on Numpy, Pandas, Sklearn, Matplotlib, and PyTorch.

\begin{table*}[t!]
\caption{RQ1: Effectiveness of DSrepair against the baselines. Values are shown in the format ANF (FR). ANF is the Absolute Number of Fixes. FR is Fix Rate. The results indicate that DSrepair outperforms the baselines for the majority of the libraries.}
\centering
\resizebox{.9\textwidth}{!}{
\begin{tabular}{l l r r r r r r r r}
\toprule
Model & Approach & Numpy & Pandas & Scipy & Sklearn & Matplotlib & PyTorch & TensorFlow & Total\\
\midrule

\multirow{6}{*}{GPT-3.5-turbo} & Code-Search & 5 (4.63\%) & 4 (2.12\%) & 2 (3.33\%) & 11 (14.86\%) & 0 (0.00\%) & 2 (4.55\%) & 2 (7.41\%) & 26 (4.63\%) \\
& Chat-Repair & 21 (19.44\%) & 7 (3.70\%) & 5 (8.33\%) & 18 (24.32\%) & 4 (6.67\%) & 6 (13.64\%) & 2 (7.41\%) & 63 (11.21\%) \\
& Self-Debugging-S & 14 (12.96\%) & 5 (2.65\%) & 6 (10.00\%) & 13 (17.57\%) & 4 (6.67\%) & 7 (15.91\%) & 2 (7.41\%) & 51 (9.07\%) \\
& Self-Debugging-E & 20 (18.52\%) & \textbf{19 (10.05\%)} & 2 (3.33\%) & 14 (18.92\%) & 6 (10.00\%) & 3 (6.82\%) & \textbf{4 (14.81\%)} & 68 (12.10\%) \\
& Self-Repair & 17 (15.74\%) & 17 (8.99\%) & 5 (8.33\%) & 12 (16.22\%) & 8 (13.33\%) & 9 (20.45\%) & \textbf{4 (14.81\%)} & 72 (12.81\%) \\
\cmidrule{2-10}
& \textbf{DSrepair} & \textbf{24 (22.22\%)} & 17 (8.99\%) & \textbf{15 (25.00\%)} & \textbf{20 (27.03\%)} & \textbf{10 (16.67\%)} & \textbf{15 (34.09\%)} & 3 (11.11\%) & \textbf{104 (18.51\%)} \\

\midrule
\multirow{6}{*}{GPT-4o-mini} & Code-Search & 28 (25.93\%) & 21 (11.11\%) & 11 (18.33\%) & 11 (14.86\%) & 15 (25.00\%) & 14 (31.82\%) & 3 (11.11\%) & 103 (18.33\%) \\
& Chat-Repair & 29 (26.85\%) & 28 (14.81\%) & 14 (23.33\%) & 19 (25.68\%) & 14 (23.33\%) & 13 (29.55\%) & \textbf{5 (18.52\%)} & 122 (21.71\%) \\
& Self-Debugging-S & 32 (29.63\%) & 25 (13.23\%) & \textbf{16 (26.67\%)} & 20 (27.03\%) & 7 (11.67\%) & 14 (31.82\%) & 4 (14.81\%) & 118 (21.00\%) \\
& Self-Debugging-E & \textbf{35 (32.41\%)} & \textbf{33 (17.46\%)} & 12 (20.00\%) & 16 (21.62\%) & 10 (16.67\%) & 17 (38.64\%) & 4 (14.81\%) & 127 (22.60\%) \\
& Self-Repair & 34 (31.48\%) & 32 (16.93\%) & 13 (21.67\%) & 15 (20.27\%) & 11 (18.33\%) & 15 (34.09\%) & \textbf{5 (18.52\%)} & 125 (22.24\%) \\
\cmidrule{2-10}
& \textbf{DSrepair} & 33 (30.56\%) & 20 (10.58\%) & 15 (25.00\%) & \textbf{31 (41.89\%)} & \textbf{22 (36.67\%)} & \textbf{19 (43.18\%)} & \textbf{5 (18.52\%)} & \textbf{145 (25.80\%)} \\

\midrule
\multirow{6}{*}{DeepSeek-Coder} & Code-Search & 23 (21.30\%) & 11 (5.82\%) & 2 (3.33\%) & 12 (16.22\%) & 17 (28.33\%) & 8 (18.18\%) & 4 (14.81\%) & 77 (13.70\%) \\
& Chat-Repair & 39 (36.11\%) & 22 (11.64\%) & 8 (13.33\%) & 18 (24.32\%) & 20 (33.33\%) & 14 (31.82\%) & 7 (25.93\%) & 128 (22.78\%) \\
& Self-Debugging-S & 33 (30.56\%) & 26 (13.76\%) & 9 (15.00\%) & 11 (14.86\%) & 13 (21.67\%) & 10 (22.73\%) & 7 (25.93\%) & 109 (19.40\%) \\
& Self-Debugging-E & 28 (25.93\%) & 23 (12.17\%) & 3 (5.00\%) & 18 (24.32\%) & 12 (20.00\%) & 10 (22.73\%) & 6 (22.22\%) & 100 (17.79\%) \\
& Self-Repair & \textbf{40 (37.04\%)} & 22 (11.64\%) & \textbf{12 (20.00\%) }& 24 (32.43\%) & 17 (28.33\%) & 11 (25.00\%) & \textbf{10 (37.04\%)} & 136 (24.20\%) \\
\cmidrule{2-10}
& \textbf{DSrepair} & 38 (35.19\%) & \textbf{28 (14.81\%)} & 10 (16.67\%) & \textbf{31 (41.89\%)} & \textbf{23 (38.33\%)} & \textbf{24 (54.55\%)} & \textbf{10 (37.04\%)} & \textbf{164 (29.18\%)} \\

\midrule
\multirow{6}{*}{Codestral} & Code-Search & 27 (25.00\%) & 13 (6.88\%) & 9 (15.00\%) & 24 (32.43\%) & 19 (31.67\%) & 8 (18.18\%) & \textbf{6 (22.22\%)} & 106 (18.86\%) \\
& Chat-Repair & 28 (25.93\%) & 13 (6.88\%) & \textbf{12 (20.00\%)} & 21 (28.38\%) & 16 (26.67\%) & 10 (22.73\%) & 5 (18.52\%) & 105 (18.68\%) \\
& Self-Debugging-S & 27 (25.00\%) & 19 (10.05\%) & 7 (11.67\%) & 13 (17.57\%) & 8 (13.33\%) & 9 (20.45\%) & 2 (7.41\%) & 85 (15.12\%) \\
& Self-Debugging-E & 26 (24.07\%) & 21 (11.11\%) & 8 (13.33\%) & 16 (21.62\%) & 10 (16.67\%) & 11 (25.00\%) & 4 (14.81\%) & 96 (17.08\%) \\
& Self-Repair & \textbf{32 (29.63\%)} & 17 (8.99\%) & 6 (10.00\%) & 14 (18.92\%) & 10 (16.67\%) & 12 (27.27\%) & 5 (18.52\%) & 96 (17.08\%) \\
\cmidrule{2-10}
& \textbf{DSrepair} & \textbf{32 (29.63\%)} & \textbf{30 (15.87\%)} & 9 (15.00\%) & \textbf{28 (37.84\%)} & \textbf{21 (35.00\%)} & \textbf{17 (38.64\%)} & 3 (11.11\%) & \textbf{140 (24.91\%)} \\

\bottomrule
\end{tabular}
\label{table: FR all library categories}
}
\vspace*{-0.4cm}
\end{table*}

\vspace{-0.0cm}
\begin{figure}[h!]
\centerline{\includegraphics[width=0.72\linewidth]{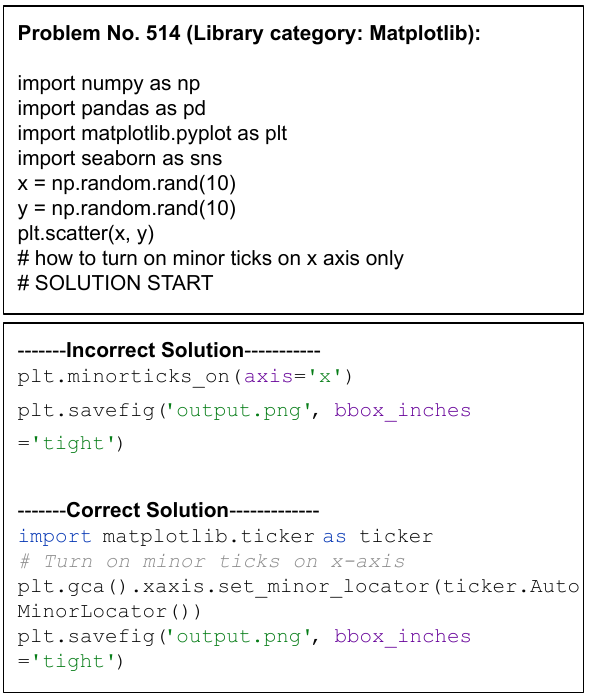}}
\caption{A code problem example from DS-1000. The incorrect solution is generated from Self-Repair, and the correct solution is generated from DSrepair. 
By incorporating knowledge of the invoked API, DSrepair can assist LLMs in generating solutions with correct API usage.
}
\label{fig: case analysis}
\vspace*{-0.2cm}
\end{figure}



Fig~\ref{fig: case analysis} shows an example from Codestral where the error can be solved by DSrepair, but cannot be solved by Self-Repair.
The purpose of this code problem is to only turn on minor ticks on the x-axis.
Self-Repair generates an incorrect fix, while DSrepair generates the correct fix.
In this problem, the buggy code uses the function \textit{plt.minorticks\_on} (short for \textit{matplotlib.pyplot.minorticks\_on}), with parameter \textit{axis=`x'}.
However, as stated in the Matplotlib official document, the full expression of \textit{plt.minorticks\_on} is \textit{matplotlib.pyplot.minorticks\_on()} with no parameters, which means that \textit{plt.minorticks\_on} can control the display of minor ticks on both x-axis and y-axis, but there is no optional parameter to control the display on x-axis or y-axis only.
With DSrepair, by enriching the prompt with knowledge of how to use \textit{plt.minorticks\_on} correctly, LLM is more likely to realize that putting parameters in function \textit{plt.minorticks\_on} is incorrect.
The correct solution uses \textit{plt.gca().xaxis.set\_minor\_locator()} instead to reach the goal of the code problem.

Looking deeper into the buggy code that DSrepair cannot fix, we identify two primary reasons. 
Firstly, the presence of multiple errors in the code poses a significant challenge.
DSrepair is designed to address specific errors highlighted by standard error messages.
However, when a code segment contains hidden bugs that come out only after fixing one bug, our \major{approach} struggles to resolve all issues in a single request.
Secondly, the insufficiency of information provided from the test cases in the description limits the repair effectiveness.
Some descriptions lack accompanying test cases, which are crucial for identifying and fixing errors.
For instance, if the buggy code triggers an assertion error, the absence of concrete test cases impedes the LLM's ability to generate a precise fix.
Even when generated code passes the given test cases, it may still fail during actual evaluation.
Simply informing the LLM that the code is incorrect without detailed guidance is often inadequate for effective repair.

\begin{tcolorbox}
\textbf{\underline{Answer to RQ1:}}
DSrepair significantly outperforms all the baselines in fixing buggy data science code.
Specifically, DSrepair demonstrates notable improvements across four LLMs by fixing 104, 145, 164, and 140 buggy programs respectively, with improvement rates of 44.4\%, 14.2\%, 20.6\%, and 32.1\%  
compared to the second-best baseline, respectively. 
\end{tcolorbox}

Please note that our comparison with baselines for detecting general software logic bugs is not meant to imply that DSrepair outperforms these baselines in all domains. Rather, we show that approaches not specifically designed for data science struggle with addressing data science bugs. By addressing the unique needs of data science code, DSrepair can significantly improve repair outcomes in the context of data science.

\subsection{RQ2: Overlap with Baseline}

\begin{figure*}[h!]
\centering
  \begin{subfigure}{0.24\linewidth}
    \hspace*{-0.7cm}
    \includegraphics[width=1.1\linewidth, trim={0 1.7cm 0 1.3cm}, clip]{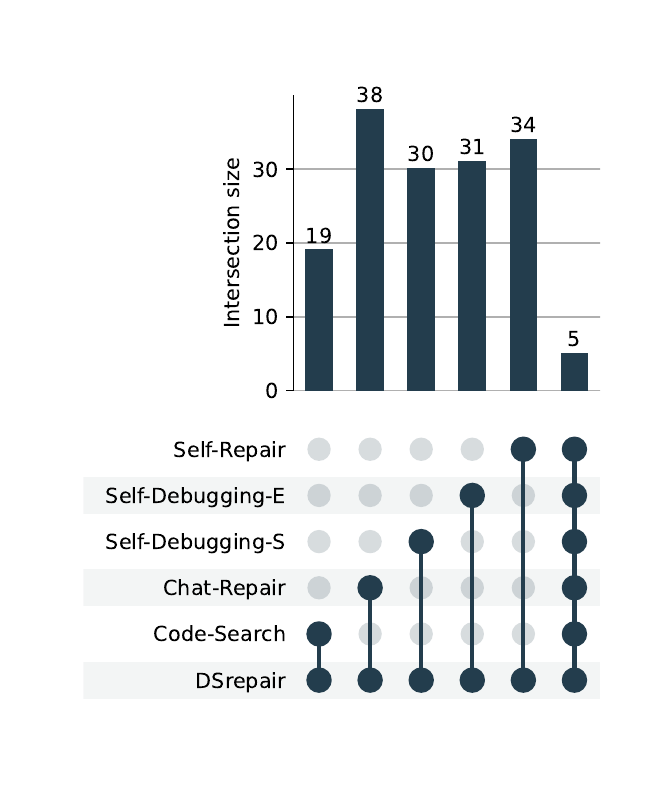}
    \caption{GPT-3.5-turbo}
  \end{subfigure}
    \begin{subfigure}{0.24\linewidth}
    \hspace*{-0.7cm}
    \includegraphics[width=1.1\linewidth, trim={0 1.7cm 0 1.3cm}, clip]{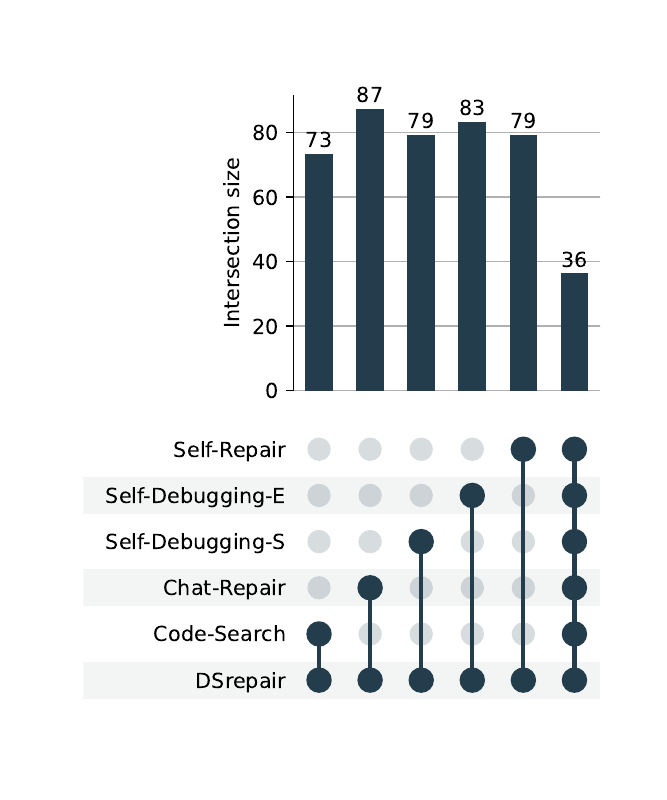}
    \caption{GPT-4o-mini}
  \end{subfigure}
  \begin{subfigure}{0.24\linewidth}
    \hspace*{-0.7cm}
    \includegraphics[width=1.1\linewidth, trim={0 1.7cm 0 1.3cm}, clip]{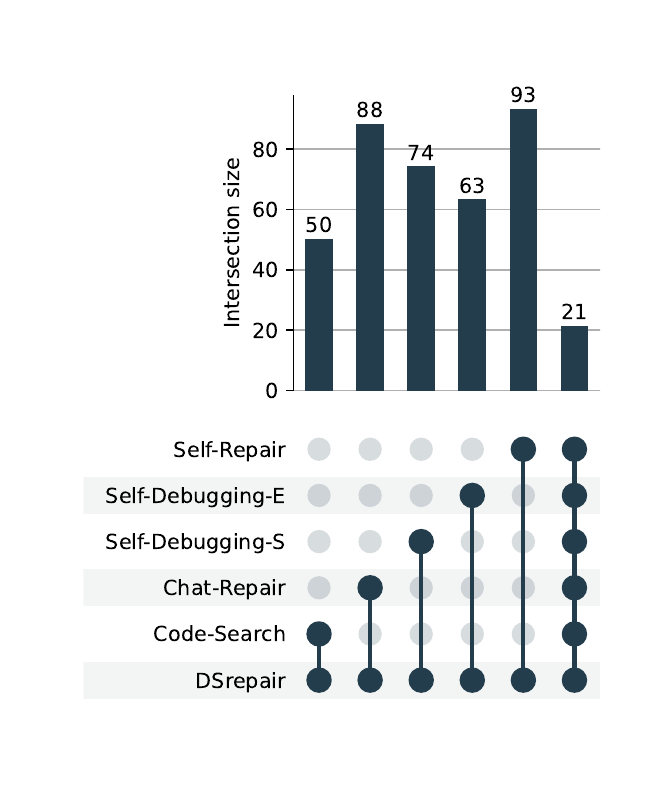}
    \caption{DeepSeek-Coder}
  \end{subfigure}
  \begin{subfigure}{0.24\linewidth}
      \hspace*{-0.7cm}
    \includegraphics[width=1.1\linewidth, trim={0 1.7cm 0 1.3cm}, clip]{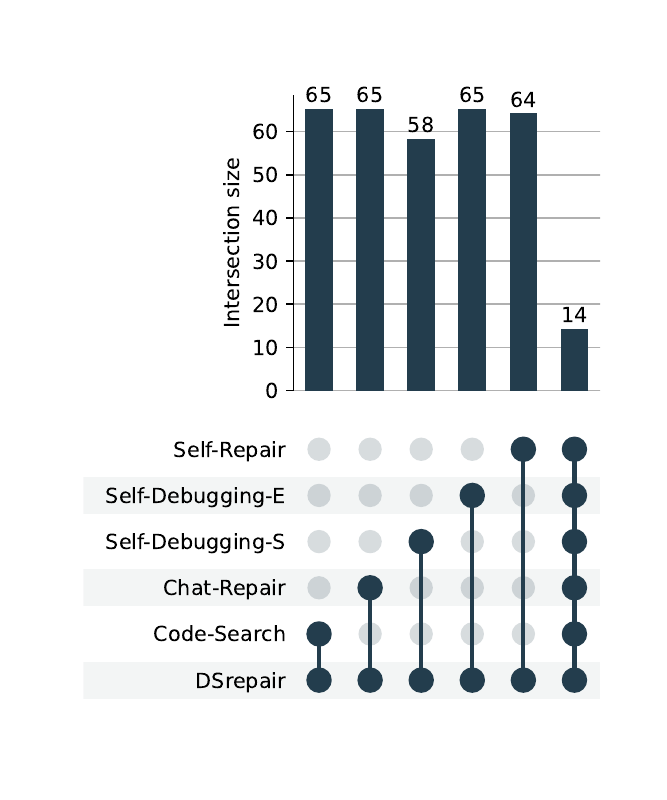}
    \caption{Codestral}
  \end{subfigure}
\caption{RQ2: Upset plots for overlap analysis.
For example, in (a), the first column indicates that
19 buggy code snippets that can be fixed by both DSrepair and Code-Search.}
\label{fig: RQ2 upsetplot}
\end{figure*}

In this RQ, we conduct an overlap analysis by comparing the solved buggy code snippets between DSrepair and the baselines.
Fig~\ref{fig: RQ2 upsetplot} shows the upset plots\cite{upsetplot} for different \major{approaches} and the intersection of their ANF.

We can observe that the fixed buggy code overlaps between DSrepair and \major{baselines} are overall less than 55\% of the bug fixes from DSrepair.
This means that about half of the code fixes from DSrepair could not be fixed by \major{baselines}.
For example, in Fig~\ref{fig: RQ2 upsetplot}(a), DSrepair can fix 104 buggy code snippets, while Self-Repair can only fix 72 buggy code snippets.
The overlap between their fixed code snippets is only 34, which means that DSrepair has 70 (104-34=70) code snippets that Self-Repair cannot fix, and Self-Repair has 38 (72-34=38) code snippets that DSrepair cannot fix.
The overlap among the six \major{baselines} is quite low (5 for GPT-3.5-turbo, 36 for GPT-4o-mini, 21 for DeepSeek-Coder, and 14 for Codestral).
Overall, each \major{baseline} shows the uniqueness of buggy code repair.



\begin{tcolorbox}
\textbf{\underline{Answer to RQ2:}}
DSrepair uniquely fixes approximately 55\% of buggy code snippets that \major{baselines} are unable to fix. 
\end{tcolorbox}

\subsection{RQ3: Cost of DSrepair}

To answer RQ3, we assess the financial costs associated with using DSrepair by quantifying the US dollar spent on interactions with using the APIs of the four LLMs.
The cost of each request to these models depends directly on the number of tokens processed, including both the tokens used for input and those generated as output.
We calculate the expenses incurred during these interactions by measuring the Token Usage (TU) of DSrepair and then converting this usage into actual Money Spent (MS), comparing these against the cost of our baselines. 

\begin{figure}[htbp]
\vspace*{-0.4cm}
\centerline{\includegraphics[width=0.8\linewidth]{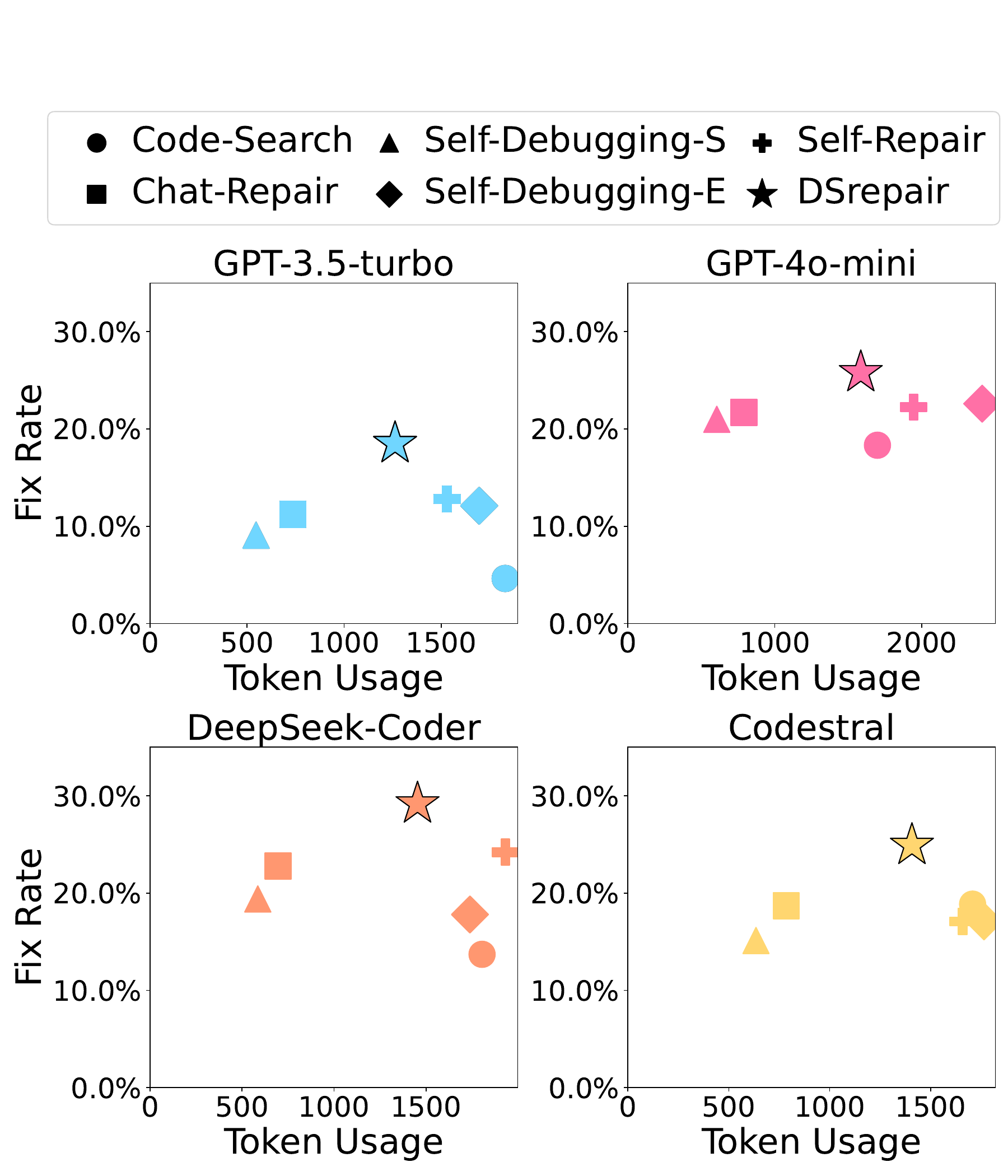}}
\caption{RQ3: Scatter plot of TU (Token Usage) and FR (Fix Rate).
DSrepair is the optimal approach (the star markers) compared with baselines.}
\label{fig: RQ2 scatter plot}
\vspace*{-0.4cm}
\end{figure}

Table~\ref{table: token usage} shows the TU and MS for different \major{approaches}.
Fig~\ref{fig: RQ2 scatter plot} shows a scatter plot of TU and FR.
We observe that DSrepair costs less token usage than the second-best baseline.
For example, 
DSrepair uses only 1262.14, 1584.74, 1453.96, and 1407.15 tokens per code problem, while Self-Repair needs 1529.63, 1944.56, 1931.20, and 1657.99 tokens per code problem.
Based on the real-time price in Table~\ref{tab: LLM detail}, the money spent on each request is \$0.00073, \$0.00043,  \$0.00025, and \$0.00185 for using GPT-3.5-turbo, GPT-4o-mini, DeepSeek-Coder, and Codestral as LLM respectively.

We can also observe that DSrepair's token usage with GPT-4o-mini, DeepSeek-Coder, and Codestral is higher than when using GPT-3.5-turbo.
This is because the return of these code LLMs may not follow the prompt's output format instruction.
The responses typically contain more information, such as line-by-line code comments, natural language explanation of the code, and an analysis of why the first generated code is incorrect, all of which contribute to extra costs.





\begin{table*}[h!]\scriptsize
\caption{RQ3: Cost of different approaches. TU refers to Token Usage (input token usage + output token usage), and MS refers to Money Spent for LLM receiving the prompt and generating the response.}
\vspace{0mm}
\centering
\begin{tabular}{l r r r r r r r r}
\toprule

\multirow{2}{*}{Approach} & \multicolumn{2}{c}{GPT-3.5-turbo} & \multicolumn{2}{c}{GPT-4o-mini} & \multicolumn{2}{c}{DeepSeek-Coder} & \multicolumn{2}{c}{Codestral} \\
\cmidrule{2-9}
 & TU & MS & TU & MS & TU & MS & TU & MS\\
\midrule
Code-Search & 1829.66 & \$0.00124 & 1697.94 & \$0.00034 & 1804.85 & \$0.00030 & 1707.45 & \$0.00218  \\
Chat-Repair & 736.22 & \$0.00050 & 788.97 & \$0.00022 & 696.66 & \$0.00012 & 784.52 & \$0.00121  \\
Self-Debugging-S & \textbf{546.13} & \textbf{\$0.00041} & \textbf{605.48} & \textbf{\$0.00018} & \textbf{585.90} & \textbf{\$0.00011} & \textbf{634.84} & \textbf{\$0.00108} \\
Self-Debugging-E & 1695.55 & \$0.00120 & 2410.03 & \$0.00070 & 1738.98 & \$0.0003 & 1763.98 & \$0.00260 \\
Self-Repair & 1529.63 & \$0.00104 & 1944.56 & \$0.00053 & 1931.20 & \$0.00034 & 1657.99 & \$0.00232 \\
\midrule
DSrepair & 1262.14 & \$0.00073 & 1584.74 & \$0.00043 & 1453.96 & \$0.00025 & 1407.15 & \$0.00185  \\

\bottomrule
\end{tabular}
\label{table: token usage}
\vspace*{-0.4cm}
\end{table*}


\begin{tcolorbox}
\textbf{\underline{Answer to RQ3:}}
Compared to the second-best baseline, DSrepair uses fewer tokens (1262.14, 1584.74, 1453.96, and 1407.15), saving 17.49\%, 34.24\%, 24.71\%, and 17.59\% tokens per code task respectively across different LLMs.
\end{tcolorbox}

\subsection{RQ4: Influence of Prompt Design}
To figure out how different prompt components influence DSrepair, we conduct an ablation study.
In DSrepair, there are two key components, i.e., API knowledge and bug knowledge.
In the ablation study, we compare DSrepair's performance with the performance of `No knowledge' (prompt without API and bug knowledge), `API Knowledge only' (prompt without bug knowledge provided by tests), and `Bug Knowledge only' (prompt without API knowledge)
We use  `DSrepair w/o API\&Bug', `DSrepair w/o Bug', and `DSrepair w/o API' to represent `No knowledge', `API Knowledge only', and `Bug Knowledge only' for short. 

The results of the ablation study are shown in Table~\ref{table: Ablation Study Fix Rate}.
When using GPT-3.5-turbo as LLM, the overall performance of DSrepair (18.51\% FR) is better than DSrepair w/o Bug (14.77\% FR) and DSrepair w/o API (16.73\% FR).
The overall performance for GPT-4o-mini of DSrepair (25.80\% FR) is better than DSrepair w/o Bug (23.49\% FR) and DSrepair w/o API (22.06\% FR).
When using DeepSeek-Coder as our LLM, DSrepair still stands for the best, with 29.18\% total FR.
However, DSrepair w/o Bug has better overall performance than DSrepair w/o API, where using DSrepair w/o Bug has 28.83\% FR while using DSrepair w/o API only has 27.94\% FR.
Using Codestral as LLM, DSrepair has 24.91\% FR, which is higher than both DSrepair w/o Bug (24.51\%) and DSrepair w/o API (24.38\%).
Interestingly, we observe that the FR declines in DSrepair w/o Bug (GPT-3.5-turbo), DSrepair w/o Bug and w/o API (GPT-4o-mini) and DSrepair w/o API (DeepSeek-Coder) compared with DSrepair w/o API\&Bug.


\begin{table}[h!]\scriptsize
\caption{RQ4: Results of ablation study. `DSrepair w/o API\&Bug' is for prompt without API and bug knowledge, `DSrepair w/o Bug' is for prompt without bug knowledge, and `DSrepair w/o API' refers to prompt without API knowledge. ANF is the Absolute Number of Fixes. FR is Fix Rate.}
\centering
\vspace{0mm}
\resizebox{.38\textwidth}{!}{

\begin{tabular}{l l r r}
\toprule
Model & Prompt & ANF & FR\\
\midrule

\multirow{4}{*}{GPT-3.5-turbo} & DSrepair w/o API & 94 & 16.73\% \\
& DSrepair w/o Bug & 83 & 14.77\% \\
& DSrepair w/o API\&Bug & 85 & 15.12\% \\
\cmidrule{2-4}
& \textbf{DSrepair} & \textbf{104} & \textbf{18.51\%} \\

\midrule
\multirow{4}{*}{GPT-4o-mini} & DSrepair w/o API & 124 & 22.06\% \\
& DSrepair w/o Bug & 132 & 23.49\% \\
& DSrepair w/o API\&Bug & 133 & 23.67\% \\
\cmidrule{2-4}
& \textbf{DSrepair} & \textbf{145} & \textbf{25.80\%} \\

\midrule
\multirow{4}{*}{DeepSeek-Coder} & DSrepair w/o API & 157 & 27.94\% \\
& DSrepair w/o Bug & 162 & 28.83\% \\
& DSrepair w/o API\&Bug & 160 & 28.47\% \\
\cmidrule{2-4}
& \textbf{DSrepair} & \textbf{164} & \textbf{29.18\%} \\

\midrule
\multirow{4}{*}{Codestral} & DSrepair w/o API & 137 & 24.38\% \\
& DSrepair w/o Bug & 139 & 24.51\% \\
& DSrepair w/o API\&Bug & 123 & 21.89\% \\
\cmidrule{2-4}
& \textbf{DSrepair} & \textbf{140} & \textbf{24.91\%} \\


\bottomrule
\end{tabular}
\label{table: Ablation Study Fix Rate}
}
\vspace*{-0.4cm}
\end{table}

\begin{tcolorbox}
\textbf{\underline{Answer to RQ4:}}
Both enriched API knowledge and enriched bug knowledge in the prompt contribute to the final effectiveness of DSrepair.
\end{tcolorbox}

\subsection{RQ5: Comparison of Different Knowledge Retrieval Approaches}

To answer RQ5, we examine the impact of various knowledge retrieval \major{approaches} on the performance of DSrepair.
Specifically, we compare knowledge retrieval through KG (DSrepair) with knowledge retrieval through plain-text searching.
For plain-text searching, we extract API knowledge using invoked API names as keywords.
The API knowledge is retrieved as a window of text encompassing 50 tokens per keyword.
This window length was chosen to match the average size of the retrieval results from DS-KG for each keyword, ensuring a fair comparison.
All other experimental settings are kept consistent with those used in DSrepair.

Table~\ref{table: plain text} shows the results of different knowledge retrieval approaches for DSrepair.
We can see that retrieving knowledge from plain text only has 14.77\%, 22.60\%, 25.44\%, and 22.78\% Fix Rate for four tested LLMs.
Retrieving knowledge from plain text uses 1432.42, 1827.59, 1723.77, and 1636.46 tokens per buggy code, which is higher than retrieval from DS-KG, and thus has higher Money Spent on four LLMs.

\begin{tcolorbox}
\textbf{\underline{Answer to RQ5:}}
Knowledge graph-based retrieval outperforms plain text-based retrieval in fixing buggy data science code.
The former's fix rate is 18.51\%, 25.80\%, 29.18\%, and 24.91\% for GPT-3.5-turbo, GPT-4o-mini, DeepSeek-Coder, and Codestral, respectively, compared to 14.77\%, 22.60\%, 25.44\%, and 22.78\% for the latter.  
\end{tcolorbox}




\begin{table}[h!]\scriptsize
\caption{RQ5: Knowledge retrieval comparison between plain text and KG. Retrieval from KG is better than from plain text. FR refers to Fix Rate, TU refers to Token Usage (input token usage + output token usage), and MS refers to Money Spent for LLM receiving the prompt and generating the response.}
\vspace{0mm}
\centering
\begin{tabular}{l l r r r}
\toprule
Model & Knowledge Retrieval & FR & TU & MS\\
\midrule

\multirow{2}{*}{GPT-3.5-turbo} & Plain Text & 14.77\% & 1432.42 & \$0.00082 \\

 & Knowledge Graph & \textbf{18.51\%} & \textbf{1262.14} & \textbf{\$0.00073} \\
\midrule
\multirow{2}{*}{GPT-4o-mini} & Plain Text & 22.60\% & 1827.59 & \$0.00046 \\
 & Knowledge Graph & \textbf{25.80\%} & \textbf{1584.74} & \textbf{\$0.00043} \\
 \midrule
\multirow{2}{*}{DeepSeek-Coder} & Plain Text & 25.44\% & 1723.77 & \$0.00030 \\
 & Knowledge Graph & \textbf{29.18\%} & \textbf{1453.96} & \textbf{\$0.00025} \\
 \midrule
\multirow{2}{*}{Codestral} & Plain Text & 22.78\% & 1636.46 & \$0.00209 \\
 & Knowledge Graph & \textbf{24.91\%} & \textbf{1407.15} & \textbf{\$0.00185} \\


\bottomrule
\end{tabular}
\label{table: plain text}
\vspace*{-0.5cm}
\end{table}

\subsection{RQ6: Influence of API Richness}
\label{section: API richness}

To address RQ6, we assess how varying the richness of API knowledge impacts the performance of DSrepair.
In our DSrepair setup, we use only the full expressions of the invoked API to enrich the prompts.
Our queries also yield additional information about correct API usage, including explanations of functions, and details about parameters and returns.
To explore the potential benefits of this enriched API knowledge, we design experiments with different richness levels of API information: DSrepair+explanation, DSrepair+parameter\&return, and DSrepair+explanation+parameter\&return.
DSrepair+explanation incorporates explanations of the invoked API into the API knowledge.
DSrepair+parameter\&return adds information about the function's parameters and returns.
DSrepair+explanation+parameter\&return combines both types of information into the API knowledge. 

Table~\ref{table: richness of API knowledge} presents the performance results of these different levels of API knowledge richness.
The results are evaluated in terms of effectiveness, as measured by the Fix Rate (FR), and cost, as quantified by Token Usage (TU) and Money Spent (MS).
The data shows that DSrepair achieves the highest Fix Rate across all richness levels, with 18.51\% on GPT-3.5-turbo, 25.80\% on GPT-4o-mini, 29.18\% on DeepSeek-Coder, and 24.91\% on Codestral.
This suggests that the additional information may complicate the prompt without necessarily improving the effectiveness of the repair.

In terms of cost, DSrepair generally exhibits lower token usage and monetary cost compared to its enriched counterparts.
For example, on GPT-3.5-turbo, DSrepair uses 1262.14 tokens and incurs a cost of \$0.00073, whereas DSrepair+explanation+parameter\&return uses 1584.49 tokens and costs \$0.00089.
This pattern holds across the other models as well, indicating that increasing the complexity of the API knowledge may lead to higher money costs without a proportional gain in repair effectiveness.


\begin{tcolorbox}
\textbf{\underline{Answer to RQ6:}}
Using full expressions of invoked API from the retrieval results in DSrepair performs the best in fixing bugs.
\end{tcolorbox}

\begin{table*}[h!]\scriptsize
\caption{RQ6: Influence of API knowledge richness on DSrepair. FR refers to Fix Rate, TU refers to Token Usage (input token usage + output token usage), and MS refers to Money Spent for LLM receiving the prompt and generating the response.}
\vspace{0mm}
\setlength{\tabcolsep}{4pt}
\centering
\begin{tabular}{l r r r r r r r r r r r r}
\toprule

\multirow{2}{*}{API Knowledge Richness} & \multicolumn{3}{c}{GPT-3.5-turbo} & \multicolumn{3}{c}{GPT-4o-mini} & \multicolumn{3}{c}{DeepSeek-Coder} & \multicolumn{3}{c}{Codestral} \\
\cmidrule{2-13}
 & FR & TU & MS & FR & TU & MS & FR & TU & MS & FR & TU & MS \\
\midrule

DSrepair+explanation & 15.84\% & 1279.50 & \$0.00074 & 22.78\% & 1597.57 & \$0.00043 & 27.22\% & 1503.78 & \$0.00026 & 24.38\% & 1410.30 & \$0.00187  \\
DSrepair+parameter\&return & 14.95\% & 1573.98 & \$0.00089 & 22.60\% & 1885.12 & \$0.00047 & 26.33\% & 1803.51 & \$0.00030 & 24.02\% & 1698.39 & \$0.00215  \\
DSrepair+explanation+parameter\&return & 15.30\% & 1584.49 & \$0.00089 & 23.49\% & 1901.94 & \$0.00047 & 26.87\% & 1816.90 & \$0.00030 & 24.02\% & 1712.37 & \$0.00217  \\
\midrule
\textbf{DSrepair} & \textbf{18.51\%} & \textbf{1262.14} & \textbf{\$0.00073} & \textbf{25.80\%} & \textbf{1584.74} & \textbf{\$0.00043} & \textbf{29.18\%} & \textbf{1453.96} & \textbf{\$0.00025} & \textbf{24.91\%} & \textbf{1407.15} & \textbf{\$0.00185}  \\


\bottomrule
\end{tabular}
\label{table: richness of API knowledge}
\end{table*}



\subsection{RQ7: Influence of LLM Non-determinism}
\label{section: RQ7}

To investigate RQ7, we explore the effect of non-determinism in LLMs on our experimental results.
As outlined in Section \ref{section: experimental design}, we conduct each experiment ten times to account for variability in LLM responses. 
From these iterations, we select the median performance result as our final data point for analysis.
To further understand how LLM non-determinism might influence our results, we calculate the mean and standard deviation of the results across the ten trials.


Table~\ref{table: non-determinism} shows the mean FR (Fix Rate) and the standard deviation of code repair results.
We observe that the standard deviations of DSrepair are not big, which indicates the stability of our experiment results.
Additionally, with the standard deviation, DSrepair still outperforms the \major{baselines} in its mean FR, with 101.80 $\pm$ 6.71, 142.90 $\pm$ 6.44, 163.67 $\pm$ 1.25, 137.80 $\pm$ 4.60 for using GPT-3.5-turbo, GPT-4o-mini, DeepSeek-Coder, and Codestral as LLM respectively.

\begin{table*}[h!]\scriptsize
\caption{RQ7: Mean and standard deviation of the ANF (Absolute Number of Fix), expressed as mean ANF $\pm$ standard deviation. The small standard deviation indicates the reliability of our results, and our experimental conclusions in RQ1 remain valid within this range.
}
\vspace{0mm}
\hspace*{-0.4cm}

\centering
\begin{tabular}{l r r r r}
\toprule
\multirow{1}{*}{Approach} & \multicolumn{1}{c}{GPT-3.5-turbo} & \multicolumn{1}{c}{GPT-4o-mini} & \multicolumn{1}{c}{DeepSeek-Coder} & \multicolumn{1}{c}{Codestral}\\
\midrule
Code-Search & 26.40 $\pm$ 2.46 & 102.5 $\pm$ 2.29 & 75.33 $\pm$ 3.09 & 104.3 $\pm$ 2.83 \\
Chat-Repair & 60.50 $\pm$ 3.75 & 122.0 $\pm$ 2.79 & 128.00 $\pm$ 0.82 & 104.8 $\pm$ 2.18 \\
Self-Debugging-S & 51.10 $\pm$ 4.44 & 117.4 $\pm$ 2.69 & 108.00 $\pm$ 2.16 & 81.70 $\pm$ 6.39 \\
Self-Debugging-E & 68.60 $\pm$ 6.18 & 124.2 $\pm$ 6.16 & 102.67 $\pm$ 3.77 & 94.60 $\pm$ 5.70 \\
Self-Repair & 71.20 $\pm$ 4.58 & 125.1 $\pm$ 3.59 & 134.33 $\pm$ 3.09 & 91.50 $\pm$ 8.42 \\
\midrule
\textbf{DSrepair} & \textbf{101.80 $\pm$ 6.71} & \textbf{142.90 $\pm$ 6.44} & \textbf{163.67 $\pm$ 1.25} & \textbf{137.80 $\pm$ 4.60} \\

\bottomrule
\end{tabular}
\label{table: non-determinism}
\end{table*}

\begin{tcolorbox}
\textbf{\underline{Answer to RQ7:}}
Despite the randomness of LLMs, DSrepair consistently outperforms the baselines with greater stability across multiple trials. It achieves mean Fix Rates of 101.80 $\pm$ 6.71, 142.90 $\pm$ 6.44, 163.67 $\pm$ 1.25, 137.80 $\pm$ 4.60 across GPT-3.5-turbo, GPT-4o-mini, DeepSeek-Coder, and Codestral respectively.
\end{tcolorbox}

\section{Discussion}
\label{section: discussion}

In this section, we discuss the threats to validity, the limitations, and the generalizability of our research.



\subsection{Threats to Validity}

The threats to \textit{internal} validity mainly lie in the implementation of our prompt design.
To reduce this threat, we design DSrepair with the idea of `Structuring Prompts', adapted from a handy template for structuring prompts, called CO-STAR framework ~\cite{promptcompetition}.
Considering key aspects that influence the effectiveness and relevance of an LLM’s response, DSrepair can lead LLMs to generate more optimal responses for code purposes.
In addition, we design research questions, such as RQ4 and RQ6, to study the influence of the different prompts on our final performance.

The threats to \textit{external} validity mainly lie in the datasets, and LLMs used in our study.
To reduce the threat regarding datasets, we carefully choose to use DS-1000 as our experiment dataset, 
which is the state-of-the-art benchmark tailored to address data leakage concerns with realistic and diverse data science problems,
with testing methods checking both execution semantics and surface-form constraints~\cite{lai2023ds}.
To reduce the threat regarding LLMs, we use four widely studied LLMs to mitigate the potential bias that certain LLMs can bring to the experiment results.
In addition, to mitigate the inherent randomness of LLMs, we experiment ten times for each \major{approach} and choose one with the median overall performance as our final result, which could further mitigate the non-determinism of LLMs. 
Moreover, we exclusively design RQ7 to study whether the non-determinism of LLMs will affect our experiment findings.

\subsection{Limitation}

The effectiveness of DSrepair depends largely on the quality and completeness of the knowledge it provides.
Our approach demonstrates capability in addressing runtime errors by eliminating the initial error.
However, this repair process can sometimes introduce or trigger new errors.
This phenomenon is particularly evident when the repaired code successfully executes but subsequently results in assertion errors.
The reliance on high-quality test cases in the problem description is crucial;
in their absence, DSrepair may guide LLMs to generate code that closely mirrors the incorrect code.
This occurs because the LLMs are provided with API knowledge that can inadvertently reinforce the use of incorrect or irrelevant APIs present in the original code.
Despite this, there are instances where, upon receiving API knowledge, the LLMs deviate from the incorrect APIs, opting instead for alternative solutions, such as using different APIs or defining new functions.

Maintaining the DS-KG presents significant challenges.
Our DS-KG only reflects the correct knowledge of APIs based on a specific version.
The rapid pace at which online API documentation is updated complicates the task of ensuring the DS-KG remains up-to-date.
Consequently, keeping the DS-KG up-to-date demands substantial effort and resources.
This maintenance burden is a critical consideration, as outdated or incomplete knowledge can adversely affect the accuracy and reliability of the repairs generated by DSrepair.
With the assistance of API document's release notes, we could manage the updating of DS-KG by leveraging library development logs to automate the process.
These logs often document changes and updates made to API libraries, allowing us to efficiently identify and integrate the necessary modifications into the KG. 

Another notable limitation of DSrepair is the time cost associated with knowledge retrieval.
When compared to plain text searching, retrieval using the DS-KG incurs a significant time overhead, averaging 51.49\% more time (approximately 0.06 seconds) per task.
While this increase in retrieval time may seem marginal, it can accumulate and impact the overall efficiency of the repair process, particularly in scenarios requiring rapid iteration and testing.

\subsection{Generalizability of DSrepair}

In this paper, DSrepair is specifically designed to enhance the repair of data science code.
Nevertheless, DSrepair's underlying methodology—leveraging knowledge-enhanced retrieval and structured bug information—can be generalized to broader coding tasks. The key innovation of DSrepair lies in its use of knowledge graph-based Retrieval-Augmented Generation (RAG), which is not inherently limited to data science APIs. By replacing the domain-specific DS-KG with a knowledge graph covering general-purpose programming languages and software libraries, DSrepair could be adapted to repair a wide range of buggy code across different domains, as well as to improve other coding tasks other than repair, such as bug localization and code generation. 

DSrepair also has the potential to be extended to support project-level code generation and repair, where understanding dependencies across multiple files and context knowledge with the same project are crucial. 
By constructing a project-specific KG that encodes function definitions, module dependencies, and code architecture, DSrepair can enhance code generation and repair in large-scale software development.

\section{Related Work}
\label{section: related work}

\subsection{Code Repair} 

The goal of automated program repair is to automatically identify and fix bugs or defects in the software.
Leveraging LLMs, such as BERT~\cite{zhang2024appt}, CodeBERT~\cite{le2023invalidator}, Codex~\cite{fan2023automated,jin2023inferfix,wu2023effective}, and GPT-series~\cite{lajko2022towards,xia2023keep,xia2023conversational,sobania2023analysis}, for code repair can achieve promising performance in generating patches for various kinds of bugs and defects.
These models are adept at grasping the core meaning and relationships within code, resulting in the generation of precise and functional fixes without the need for compilation.
Using LLMs for fixing code speeds up the identification and resolution of bugs, freeing software developers to tackle more intricate issues.
This contributes to improved software reliability and upkeep.
ChatGPT, in particular, stands out among LLMs because of its built-in interactive nature, which fosters an ongoing loop of feedback, producing patches that are more polished and appropriate to the context~\cite{xia2023keep,xia2023conversational}.

\subsection{Prompt Engineering}

Prompt designing is an increasingly important skill set needed to leverage effectively with LLMs \cite{white2023prompt}, such as ChatGPT.
Similar to software design \cite{gamma1995design}, the design of prompt aims at offering reusable solutions to specific problems, 
by providing a codified approach to customizing the output and interactions of LLMs.
Abukhalaf et al.~\cite{abukhalaf2023codex} conduct an empirical study on Object Constraint Language based constraint generation, by comparing the Codex generated constraints and humane-written constraints.
Xia et al.~\cite{xia2023conversational} specifically examined prompts for automatic code repair.
More specifically, White et al. \cite{white2023chatgpt} focus on combatting mistakes and improving generated code quality by designing prompt patterns.
Borji et al.~\cite{borji2023categorical} examine the quality of generated answers and code from LLMs, and conclude the existing failures from the experiment. 
Our research work draws inspiration from these explorations and prompts that could be used to generate code candidates with better quality and fewer errors.

\subsection{Retrieval-Augmented Generation}


RAG aims to address the limitations of generative models, including issues related to outdated knowledge, a deficiency in long-tail knowledge~\cite{mallen2022not}, and the potential for private training data leakage~\cite{carlini2021extracting}.
Early research in code generation concentrated on code-to-code retrieval using dual encoder models, with the retrieved outputs subsequently inputted into autoregressive language models~\cite{lu2022reacc}. 
RepoCoder~\cite{zhang2023repocoder} enhances retrieval processes by employing iterative incremental generations~\cite{chu2011contextual}.
KNM~\cite{tang2023domain} leverages in-domain code databases and applies Bayesian inference to finalize the generated code.
RAG also can be used to build prompts for transformer-based generative models with retrieved information, including similar examples~\cite{parvez2021retrieval, li2023acecoder}, relevant API details~\cite{zan2022language, zan2023private}, documentations~\cite{zhou2022docprompting}, and imports~\cite{liu2023codegen4libs}.

\section{Conclusion}
\label{section: conclusion}
We propose DSrepair, a novel knowledge-enhanced approach for data science code repair.
We perform experiments with four LLMs and five
baselines in data science code repair and find that DSrepair significantly outperforms all the baselines in repairing data science code.
By integrating API knowledge retrieval and bug information enrichment, we can guarantee better performance in code repair, and gain people's trust in using LLMs for coding.
In future work, 
we also plan to explore a multi-agent framework with interactive feedback to enhance the performance of DSrepair while focusing on optimizing feedback steps and resource use to ensure scalability, cost efficiency and robust data science code repair.

\section{Acknowledgement}
This work was supported by the UKRI Centre for Doctoral Training in Safe and Trusted Artificial Intelligence (EP/S023356/1), the NSFC (62192732), and the National Natural Science Foundation of China (62402482).


\bibliographystyle{IEEEtran}
\bibliography{reference}

\begin{thebibliography}{10}
\providecommand{\url}[1]{#1}
\csname url@samestyle\endcsname
\providecommand{\newblock}{\relax}
\providecommand{\bibinfo}[2]{#2}
\providecommand{\BIBentrySTDinterwordspacing}{\spaceskip=0pt\relax}
\providecommand{\BIBentryALTinterwordstretchfactor}{4}
\providecommand{\BIBentryALTinterwordspacing}{\spaceskip=\fontdimen2\font plus
\BIBentryALTinterwordstretchfactor\fontdimen3\font minus \fontdimen4\font\relax}
\providecommand{\BIBforeignlanguage}[2]{{%
\expandafter\ifx\csname l@#1\endcsname\relax
\typeout{** WARNING: IEEEtran.bst: No hyphenation pattern has been}%
\typeout{** loaded for the language `#1'. Using the pattern for}%
\typeout{** the default language instead.}%
\else
\language=\csname l@#1\endcsname
\fi
#2}}
\providecommand{\BIBdecl}{\relax}
\BIBdecl

\bibitem{bolyen2019reproducible}
E.~Bolyen, J.~R. Rideout, M.~R. Dillon, N.~A. Bokulich, C.~C. Abnet, G.~A. Al-Ghalith, H.~Alexander, E.~J. Alm, M.~Arumugam, F.~Asnicar \emph{et~al.}, ``Reproducible, interactive, scalable and extensible microbiome data science using qiime 2,'' \emph{Nature biotechnology}, vol.~37, no.~8, pp. 852--857, 2019.

\bibitem{hassani2023role}
H.~Hassani and E.~S. Silva, ``The role of chatgpt in data science: how ai-assisted conversational interfaces are revolutionizing the field,'' \emph{Big data and cognitive computing}, vol.~7, no.~2, p.~62, 2023.

\bibitem{hong2024data}
S.~Hong, Y.~Lin, B.~Liu, B.~Wu, D.~Li, J.~Chen, J.~Zhang, J.~Wang, L.~Zhang, M.~Zhuge \emph{et~al.}, ``Data interpreter: An llm agent for data science,'' \emph{arXiv preprint arXiv:2402.18679}, 2024.

\bibitem{nejjar2023llms}
M.~Nejjar, L.~Zacharias, F.~Stiehle, and I.~Weber, ``Llms for science: Usage for code generation and data analysis,'' \emph{arXiv preprint arXiv:2311.16733}, 2023.

\bibitem{lai2023ds}
Y.~Lai, C.~Li, Y.~Wang, T.~Zhang, R.~Zhong, L.~Zettlemoyer, W.-t. Yih, D.~Fried, S.~Wang, and T.~Yu, ``Ds-1000: A natural and reliable benchmark for data science code generation,'' in \emph{International Conference on Machine Learning}.\hskip 1em plus 0.5em minus 0.4em\relax PMLR, 2023, pp. 18\,319--18\,345.

\bibitem{zan2023private}
D.~Zan, B.~Chen, Y.~Gong, J.~Cao, F.~Zhang, B.~Wu, B.~Guan, Y.~Yin, and Y.~Wang, ``Private-library-oriented code generation with large language models,'' \emph{arXiv preprint arXiv:2307.15370}, 2023.

\bibitem{zan2022language}
D.~Zan, B.~Chen, Z.~Lin, B.~Guan, Y.~Wang, and J.-G. Lou, ``When language model meets private library,'' \emph{arXiv preprint arXiv:2210.17236}, 2022.

\bibitem{ge2024openagi}
Y.~Ge, W.~Hua, K.~Mei, J.~Tan, S.~Xu, Z.~Li, Y.~Zhang \emph{et~al.}, ``Openagi: When llm meets domain experts,'' \emph{Advances in Neural Information Processing Systems}, vol.~36, 2024.

\bibitem{xia2023keep}
C.~S. Xia and L.~Zhang, ``Keep the conversation going: Fixing 162 out of 337 bugs for \$0.42 each using chatgpt,'' \emph{arXiv preprint arXiv:2304.00385}, 2023.

\bibitem{gupta2020synthesize}
K.~Gupta, P.~E. Christensen, X.~Chen, and D.~Song, ``Synthesize, execute and debug: Learning to repair for neural program synthesis,'' \emph{Advances in Neural Information Processing Systems}, vol.~33, pp. 17\,685--17\,695, 2020.

\bibitem{chen2023teaching}
X.~Chen, M.~Lin, N.~Sch{\"a}rli, and D.~Zhou, ``Teaching large language models to self-debug,'' \emph{arXiv preprint arXiv:2304.05128}, 2023.

\bibitem{fu2023chatgpt}
M.~Fu, C.~K. Tantithamthavorn, V.~Nguyen, and T.~Le, ``Chatgpt for vulnerability detection, classification, and repair: How far are we?'' in \emph{2023 30th Asia-Pacific Software Engineering Conference (APSEC)}.\hskip 1em plus 0.5em minus 0.4em\relax IEEE, 2023, pp. 632--636.

\bibitem{zhang2023critical}
Q.~Zhang, T.~Zhang, J.~Zhai, C.~Fang, B.~Yu, W.~Sun, and Z.~Chen, ``A critical review of large language model on software engineering: An example from chatgpt and automated program repair,'' \emph{arXiv preprint arXiv:2310.08879}, 2023.

\bibitem{lewis2020retrieval}
P.~Lewis, E.~Perez, A.~Piktus, F.~Petroni, V.~Karpukhin, N.~Goyal, H.~K{\"u}ttler, M.~Lewis, W.-t. Yih, T.~Rockt{\"a}schel \emph{et~al.}, ``Retrieval-augmented generation for knowledge-intensive nlp tasks,'' \emph{Advances in Neural Information Processing Systems}, vol.~33, pp. 9459--9474, 2020.

\bibitem{parvez2021retrieval}
M.~R. Parvez, W.~U. Ahmad, S.~Chakraborty, B.~Ray, and K.-W. Chang, ``Retrieval augmented code generation and summarization,'' \emph{arXiv preprint arXiv:2108.11601}, 2021.

\bibitem{li2023acecoder}
J.~Li, Y.~Zhao, Y.~Li, G.~Li, and Z.~Jin, ``Acecoder: Utilizing existing code to enhance code generation,'' \emph{arXiv preprint arXiv:2303.17780}, 2023.

\bibitem{zhou2022docprompting}
S.~Zhou, U.~Alon, F.~F. Xu, Z.~Wang, Z.~Jiang, and G.~Neubig, ``Docprompting: Generating code by retrieving the docs,'' \emph{arXiv preprint arXiv:2207.05987}, 2022.

\bibitem{liu2023codegen4libs}
M.~Liu, T.~Yang, Y.~Lou, X.~Du, Y.~Wang, and X.~Peng, ``Codegen4libs: A two-stage approach for library-oriented code generation,'' in \emph{2023 38th IEEE/ACM International Conference on Automated Software Engineering (ASE)}.\hskip 1em plus 0.5em minus 0.4em\relax IEEE, 2023, pp. 434--445.

\bibitem{ahmed2024automatic}
T.~Ahmed, K.~S. Pai, P.~Devanbu, and E.~Barr, ``Automatic semantic augmentation of language model prompts (for code summarization),'' in \emph{Proceedings of the IEEE/ACM 46th International Conference on Software Engineering}, 2024, pp. 1--13.

\bibitem{lu2022reacc}
S.~Lu, N.~Duan, H.~Han, D.~Guo, S.-w. Hwang, and A.~Svyatkovskiy, ``Reacc: A retrieval-augmented code completion framework,'' \emph{arXiv preprint arXiv:2203.07722}, 2022.

\bibitem{tang2023domain}
Z.~Tang, J.~Ge, S.~Liu, T.~Zhu, T.~Xu, L.~Huang, and B.~Luo, ``Domain adaptive code completion via language models and decoupled domain databases,'' in \emph{2023 38th IEEE/ACM International Conference on Automated Software Engineering (ASE)}.\hskip 1em plus 0.5em minus 0.4em\relax IEEE, 2023, pp. 421--433.

\bibitem{zhang2023repocoder}
F.~Zhang, B.~Chen, Y.~Zhang, J.~Keung, J.~Liu, D.~Zan, Y.~Mao, J.-G. Lou, and W.~Chen, ``Repocoder: Repository-level code completion through iterative retrieval and generation,'' \emph{arXiv preprint arXiv:2303.12570}, 2023.

\bibitem{yin2018learning}
P.~Yin, B.~Deng, E.~Chen, B.~Vasilescu, and G.~Neubig, ``Learning to mine aligned code and natural language pairs from stack overflow,'' in \emph{Proceedings of the 15th international conference on mining software repositories}, 2018, pp. 476--486.

\bibitem{yin2022natural}
P.~Yin, W.-D. Li, K.~Xiao, A.~Rao, Y.~Wen, K.~Shi, J.~Howland, P.~Bailey, M.~Catasta, H.~Michalewski \emph{et~al.}, ``Natural language to code generation in interactive data science notebooks,'' \emph{arXiv preprint arXiv:2212.09248}, 2022.

\bibitem{gptmodel}
https://platform.openai.com/docs/models.

\bibitem{guo2024deepseek}
D.~Guo, Q.~Zhu, D.~Yang, Z.~Xie, K.~Dong, W.~Zhang, G.~Chen, X.~Bi, Y.~Wu, Y.~Li \emph{et~al.}, ``Deepseek-coder: When the large language model meets programming--the rise of code intelligence,'' \emph{arXiv preprint arXiv:2401.14196}, 2024.

\bibitem{codestral}
https://mistral.ai/news/codestral/.

\bibitem{homepage}
https://github.com/ShuyinOuyang/DSrepair.

\bibitem{simperl2014collaborative}
E.~Simperl and M.~Luczak-R{\"o}sch, ``Collaborative ontology engineering: a survey,'' \emph{The Knowledge Engineering Review}, vol.~29, no.~1, pp. 101--131, 2014.

\bibitem{suarez2011neon}
M.~C. Su{\'a}rez-Figueroa, A.~G{\'o}mez-P{\'e}rez, and M.~Fern{\'a}ndez-L{\'o}pez, ``The neon methodology for ontology engineering,'' in \emph{Ontology engineering in a networked world}.\hskip 1em plus 0.5em minus 0.4em\relax Springer, 2011, pp. 9--34.

\bibitem{liang2022misusehint}
Q.~Liang, Z.~Kuai, Y.~Zhang, Z.~Zhang, L.~Kuang, and L.~Zhang, ``Misusehint: A service for api misuse detection based on building knowledge graph from documentation and codebase,'' in \emph{2022 IEEE International Conference on Web Services (ICWS)}.\hskip 1em plus 0.5em minus 0.4em\relax IEEE, 2022, pp. 246--255.

\bibitem{abdelaziz2021toolkit}
I.~Abdelaziz, J.~Dolby, J.~McCusker, and K.~Srinivas, ``A toolkit for generating code knowledge graphs,'' in \emph{Proceedings of the 11th Knowledge Capture Conference}, 2021, pp. 137--144.

\bibitem{prudhommeaux2008sparql}
E.~Prudhommeaux, ``Sparql query language for rdf,'' \emph{http://www. w3. org/TR/rdf-sparql-query/}, 2008.

\bibitem{blinov2020semantic}
P.~Blinov, ``Semantic triples verbalization with generative pre-training model,'' in \emph{Proceedings of the 3rd International Workshop on Natural Language Generation from the Semantic Web (WebNLG+)}, 2020, pp. 154--158.

\bibitem{abreu2007accuracy}
R.~Abreu, P.~Zoeteweij, and A.~J. Van~Gemund, ``On the accuracy of spectrum-based fault localization,'' in \emph{Testing: Academic and industrial conference practice and research techniques-MUTATION (TAICPART-MUTATION 2007)}.\hskip 1em plus 0.5em minus 0.4em\relax IEEE, 2007, pp. 89--98.

\bibitem{papadakis2015metallaxis}
M.~Papadakis and Y.~Le~Traon, ``Metallaxis-fl: mutation-based fault localization,'' \emph{Software Testing, Verification and Reliability}, vol.~25, no. 5-7, pp. 605--628, 2015.

\bibitem{brate2022improving}
R.~Brate, M.-H. Dang, F.~Hoppe, Y.~He, A.~Mero{\~n}o-Pe{\~n}uela, and V.~Sadashivaiah, ``Improving language model predictions via prompts enriched with knowledge graphs,'' in \emph{DL4KG@ ISWC2022}, 2022.

\bibitem{agashe2019juice}
R.~Agashe, S.~Iyer, and L.~Zettlemoyer, ``Juice: A large scale distantly supervised dataset for open domain context-based code generation,'' \emph{arXiv preprint arXiv:1910.02216}, 2019.

\bibitem{exact_match}
https://huggingface.co/spaces/evaluate-metric/exact\_match.

\bibitem{post2018call}
M.~Post, ``A call for clarity in reporting bleu scores,'' \emph{arXiv preprint arXiv:1804.08771}, 2018.

\bibitem{numpy}
https://numpy.org/doc/stable/index.html.

\bibitem{pandas}
https://pandas.pydata.org/docs/index.html.

\bibitem{scipy}
https://docs.scipy.org/doc/scipy/index.html.

\bibitem{sklearn}
https://scikit-learn.org/stable/.

\bibitem{matplotlib}
https://matplotlib.org/.

\bibitem{pytorch}
https://PyTorch.org/.

\bibitem{TensorFlow}
https://www.tensorflow.org/.

\bibitem{chen2024code}
J.~Chen, X.~Hu, Z.~Li, C.~Gao, X.~Xia, and D.~Lo, ``Code search is all you need? improving code suggestions with code search,'' in \emph{Proceedings of the IEEE/ACM 46th International Conference on Software Engineering}, 2024, pp. 1--13.

\bibitem{lucene}
https://lucene.apache.org/.

\bibitem{bahrami2021pytorrent}
M.~Bahrami, N.~Shrikanth, S.~Ruangwan, L.~Liu, Y.~Mizobuchi, M.~Fukuyori, W.-P. Chen, K.~Munakata, and T.~Menzies, ``Pytorrent: A python library corpus for large-scale language models,'' \emph{arXiv preprint arXiv:2110.01710}, 2021.

\bibitem{olausson2023demystifying}
T.~X. Olausson, J.~P. Inala, C.~Wang, J.~Gao, and A.~Solar-Lezama, ``Demystifying gpt self-repair for code generation,'' \emph{arXiv preprint arXiv:2306.09896}, 2023.

\bibitem{ouyang2023llm}
S.~Ouyang, J.~M. Zhang, M.~Harman, and M.~Wang, ``Llm is like a box of chocolates: the non-determinism of chatgpt in code generation,'' \emph{arXiv preprint arXiv:2308.02828}, 2023.

\bibitem{upsetplot}
https://en.wikipedia.org/wiki/UpSet\_plot.

\bibitem{promptcompetition}
https://towardsdatascience.com/how-i-won-singapores-gpt-4-prompt-engineering-competition-34c195a93d41.

\bibitem{zhang2024appt}
Q.~Zhang, C.~Fang, W.~Sun, Y.~Liu, T.~He, X.~Hao, and Z.~Chen, ``Appt: Boosting automated patch correctness prediction via fine-tuning pre-trained models,'' \emph{IEEE Transactions on Software Engineering}, 2024.

\bibitem{le2023invalidator}
T.~Le-Cong, D.-M. Luong, X.~B.~D. Le, D.~Lo, N.-H. Tran, B.~Quang-Huy, and Q.-T. Huynh, ``Invalidator: Automated patch correctness assessment via semantic and syntactic reasoning,'' \emph{IEEE Transactions on Software Engineering}, 2023.

\bibitem{fan2023automated}
Z.~Fan, X.~Gao, M.~Mirchev, A.~Roychoudhury, and S.~H. Tan, ``Automated repair of programs from large language models,'' in \emph{2023 IEEE/ACM 45th International Conference on Software Engineering (ICSE)}.\hskip 1em plus 0.5em minus 0.4em\relax IEEE, 2023, pp. 1469--1481.

\bibitem{jin2023inferfix}
M.~Jin, S.~Shahriar, M.~Tufano, X.~Shi, S.~Lu, N.~Sundaresan, and A.~Svyatkovskiy, ``Inferfix: End-to-end program repair with llms,'' in \emph{Proceedings of the 31st ACM Joint European Software Engineering Conference and Symposium on the Foundations of Software Engineering}, 2023, pp. 1646--1656.

\bibitem{wu2023effective}
Y.~Wu, N.~Jiang, H.~V. Pham, T.~Lutellier, J.~Davis, L.~Tan, P.~Babkin, and S.~Shah, ``How effective are neural networks for fixing security vulnerabilities,'' in \emph{Proceedings of the 32nd ACM SIGSOFT International Symposium on Software Testing and Analysis}, 2023, pp. 1282--1294.

\bibitem{lajko2022towards}
M.~Lajk{\'o}, V.~Csuvik, and L.~Vid{\'a}cs, ``Towards javascript program repair with generative pre-trained transformer (gpt-2),'' in \emph{Proceedings of the Third International Workshop on Automated Program Repair}, 2022, pp. 61--68.

\bibitem{xia2023conversational}
C.~S. Xia and L.~Zhang, ``Conversational automated program repair,'' \emph{arXiv preprint arXiv:2301.13246}, 2023.

\bibitem{sobania2023analysis}
D.~Sobania, M.~Briesch, C.~Hanna, and J.~Petke, ``An analysis of the automatic bug fixing performance of chatgpt,'' in \emph{2023 IEEE/ACM International Workshop on Automated Program Repair (APR)}.\hskip 1em plus 0.5em minus 0.4em\relax IEEE, 2023, pp. 23--30.

\bibitem{white2023prompt}
J.~White, Q.~Fu, S.~Hays, M.~Sandborn, C.~Olea, H.~Gilbert, A.~Elnashar, J.~Spencer-Smith, and D.~C. Schmidt, ``A prompt pattern catalog to enhance prompt engineering with chatgpt,'' \emph{arXiv preprint arXiv:2302.11382}, 2023.

\bibitem{gamma1995design}
E.~Gamma, R.~Helm, R.~Johnson, and J.~Vlissides, \emph{Design patterns: elements of reusable object-oriented software}.\hskip 1em plus 0.5em minus 0.4em\relax Pearson Deutschland GmbH, 1995.

\bibitem{abukhalaf2023codex}
S.~Abukhalaf, M.~Hamdaqa, and F.~Khomh, ``On codex prompt engineering for ocl generation: An empirical study,'' \emph{arXiv preprint arXiv:2303.16244}, 2023.

\bibitem{white2023chatgpt}
J.~White, S.~Hays, Q.~Fu, J.~Spencer-Smith, and D.~C. Schmidt, ``Chatgpt prompt patterns for improving code quality, refactoring, requirements elicitation, and software design,'' \emph{arXiv preprint arXiv:2303.07839}, 2023.

\bibitem{borji2023categorical}
A.~Borji, ``A categorical archive of chatgpt failures,'' \emph{arXiv preprint arXiv:2302.03494}, 2023.

\bibitem{mallen2022not}
A.~Mallen, A.~Asai, V.~Zhong, R.~Das, D.~Khashabi, and H.~Hajishirzi, ``When not to trust language models: Investigating effectiveness of parametric and non-parametric memories,'' \emph{arXiv preprint arXiv:2212.10511}, 2022.

\bibitem{carlini2021extracting}
N.~Carlini, F.~Tramer, E.~Wallace, M.~Jagielski, A.~Herbert-Voss, K.~Lee, A.~Roberts, T.~Brown, D.~Song, U.~Erlingsson \emph{et~al.}, ``Extracting training data from large language models,'' in \emph{30th USENIX Security Symposium (USENIX Security 21)}, 2021, pp. 2633--2650.

\bibitem{chu2011contextual}
W.~Chu, L.~Li, L.~Reyzin, and R.~Schapire, ``Contextual bandits with linear payoff functions,'' in \emph{Proceedings of the Fourteenth International Conference on Artificial Intelligence and Statistics}.\hskip 1em plus 0.5em minus 0.4em\relax JMLR Workshop and Conference Proceedings, 2011, pp. 208--214.

\end{thebibliography}

\end{document}